\documentclass{aa} 

\usepackage{graphicx}
\usepackage{txfonts}
\usepackage{array}
\usepackage{wrapfig}
\usepackage{multirow}
\usepackage{tabularx}
\usepackage[table]{xcolor}
\usepackage{color}
\begin{document} 

\title{Spectro-imagery of an active tornado-like prominence: formation and evolution}
 
 \author{Krzysztof Barczynski \inst{1,2,3}
 \and
 Brigitte Schmieder\inst{3,4,5}
 \and
 Aaron W. Peat\inst{4}
 \and
 Nicolas Labrosse\inst{4}
 \and
 Pierre Mein\inst{3}
 \and
 Nicole Mein\inst{3}
 }

\institute{PMOD/WRC, Dorfstrasse 33, CH-7260 Davos Dorf, Switzerland\\
 \email{krzysztof.barczynski@pmodwrc.ch}
 \and
 ETH-Zurich, H\"onggerberg campus, HIT building, Z\"urich, Switzerland
 \and
 LESIA, Observatoire de Paris, Universit\'e PSL , CNRS, Sorbonne Universit\'e, Universit\'e Paris-Diderot, 5 place Jules Janssen, 92190 Meudon, France
 \and
 SUPA, School of Physics and Astronomy, University of Glasgow, Scotland
 \and
 KU Leuven, Belgium}

 \date{Received ...; accepted ...;}


 \abstract
 {The dynamical nature of fine structures in prominences remains an open issue,
 including rotating flows in tornado-prominences. While the Atmospheric Imaging Assembly (AIA) imager aboard the Solar Dynamics Observatory (SDO) allowed us to follow the global structure of a tornado-like prominence during five hours, the Interface Region Imaging Spectrograph (IRIS), and the Multi subtractive Double pass spectrograph (MSDP)
 permitted to obtain plasma diagnostics of its fine structures.
 }
 {We aim to address two questions. Is the observed plasma rotation conceptually acceptable in a flux rope 
 magnetic support configuration with dips? How is the plasma density 
 distributed in the tornado-like prominence?}
 {We calculated line-of-sight velocities and non-thermal line widths using Gaussian fitting for \ion{Mg}{ii} lines and bisector method for H$\alpha$ line. We determined the electron density from \ion{Mg}{ii} line integrated intensities and profile fitting methods using 1D NLTE radiative transfer theory models.}
 { The global structure of the prominence observed in H$\alpha$, and \ion{Mg}{ii} h and k lines fits with a magnetic field structure configuration with dips.
 Coherent Dopplershifts in red- and blue-shifted areas observed in both lines were detected along rapidly-changing vertical and horizontal structures. However, the tornado at the top of the prominence consists of multiple-fine threads with opposite flows suggesting counter streaming flows rather than rotation.
 Surprisingly we found that the electron density at the top of the prominence could be larger (10$^{11}$ cm$^{-3}$) than in the inner part of the prominence. 
 }
 %
 {We suggest that the tornado is in a formation state with cooling of hot plasma in a first phase, and following that, a phase of leakage of the formed blobs with large transverse flows of material along long loops extended away of the UV prominence top.
 The existence of such long magnetic field lines on both sides of the prominence would avoid the tornado-like prominence to really turn around its axis.}


\maketitle
%

\section{Introduction}
\label{sec:intro}
Solar prominences are dense and cool plasma structures (10$^4$ K) embedded in the hot solar corona (10$^6$ K).
In chromospheric lines such as H$\alpha$, prominences observed on the solar disk appear as darker structures than the surrounding; these structures are called filaments.
They lie over magnetic-polarity inversion lines of the radial component of the photospheric magnetic field (see \citealt{Mackay2010} for a review). 
It is accepted that prominence plasma is supported in dips of magnetic field lines either forming an arcade or a flux rope where pressure tension balances the gravitational force \citep{Aulanier1998,vanBallegooijen2004,Dudik2008}.
Multi-wavelength analysis and magnetic field extrapolations provide us the magnetic field topology of large-scale structure of solar prominences (\citet{Mackay2010}).

The formation of prominences is still an open issue.
Different mechanisms suggested by observations or theory have been proposed such as levitation \citep{Okamoto2010}, injection \citep{Magara2007} or condensation 
\citep{Mackay2010}.
The mechanism of prominence formation by condensation of plasma in the dips of the magnetic field has been well developed since the paper of \citet{Karpen2001} where they showed that heating the plasma at the feet of loops generates condensation at the top.
Many models and numerical simulations are able to reproduce fine structures of prominences by using this mechanism in 1D models and further on in 3D models. The result of these simulations mimic the observations of the Atmospheric Imaging Assembly (AIA) aboard the Solar Dynamics Observatory (SDO) \citep{Karpen2001,Luna2012,Xia2014}.
Another approach to prove the existence of dips was proposed by \citet{Gunar2015,Gunar2016,Gunar2018}. Their technique attempts to recreate a 3D visualisation of H$\alpha$ observations. It starts from a 3D magnetic model, provided by simulation or extrapolation, and adds the exchange of emission between the threads using radiative transfer. Their results consisted of the prominence as viewed from different angles, allowing them to directly retrieve all the possible shapes of the simulated prominence from any angle. 

However, all of these models are based on the existence of dips. On the other hand \citet{Claes2020} focussed on hydrodynamics, creating mini thread-like features through non-linear thermal instability which do not strictly outline magnetic field structures. This instability creates blobs which, after the phase of formation, follow magnetic field lines. This could explain the fragmentation of threads in prominences and high microturbuelence found in active prominence.
In some aspects, this can be taken into account for interpreting prominence plasma during the stage of formation.
 
The determination of plasma properties is an essential component for our understanding of solar prominences, and provides important constraints on the scenarios attempting to explain their characteristics and appearance (see \citealt{Labrosse2010} for a review). Chromospheric lines such as hydrogen and \ion{Mg}{ii}~h\&k lines give good diagnostics of prominence plasma.
 
Prominences have long been observed in H$\alpha$ emission. H$\alpha$ is an optically thin line and provides information through the prominence along the line of sight \citep{Wiik1992}. Radiative transfer codes have been developed in the 90s and provide different characteristics of the hydrogen lines \citep[Lyman, Balmer and Paschen lines -- see][]{Gouutebroze1993} which can be directly compared with observations \citep{Schmieder1991,Schmieder1999}. More recently, a new series of models in 2D configurations were used to interpret the dynamics of multiple threads observed in hydrogen lines \citep{Gunar2007,Gunar2008}.
 
 With the launch of the Interface Region Imaging Spectrograph (IRIS; \cite{DePontieu2014}) in 2013, observations of \ion{Mg}{ii} lines are now at our disposal. The resonance lines of \ion{Mg}{ii}, h (2803.5\AA) and k (2796.4\AA), present self-reversed profiles on the solar disc. Their emission along the line profile represents the physical conditions of the plasma at different temperatures in the solar atmosphere. For example, along \ion{Mg}{ii} k, the k1 minimum is formed in the lower chromosphere; the k2 peaks are formed in the middle chromosphere; and the k3 self-reversal in the upper chromosphere \citep{Leenaarts2013b, Leenaarts2013}. Nevertheless, in prominences, \ion{Mg}{ii}~h and k often exhibit single peaked line profiles \citep{Levens2016,Ruan2018}. This allows us to fit the profiles with a Gaussian profile. However, in other observed prominences, wide \ion{Mg}{ii} profiles showing two peaks have been interpreted as multiple components corresponding to multiple structures crossing the line of sight and they are different than self-reversal lines observed at the solar disk \citep{Schmieder2014,Ruan2018}. 
 
The radiative transfer behind the formation of these \ion{Mg}{ii} profiles has recently been studied by \citet{Heinzel2014,Heinzel2015,Jejcic2018,Levens2019}. Electron density and optical thickness were deduced from observations of the FWHM, integrated intensities of H$\alpha$ and \ion{Mg}{ii} in prominences \citep{Ruan2018}. The core of \ion{Mg}{ii} is optically thick and the emission comes from the surface layers of prominences while the wings are optically thinner and the emission here encapsulates that of all of the structures along the line of sight. Therefore, Dopplershifts measured in the \ion{Mg}{ii} wings could be similar to those calculated in H$\alpha$, as shown by \citet{Ruan2018}.

The use of the term \textit{tornado} to describe the shape and motion of rotating H$\alpha$ prominences was introduced by \citet{Pettit1932}. Nowadays, the high temporal and spatial resolutions of SDO/AIA can be exploited to observe tornado-like prominences. These tornadoes are reported in a few papers as helical structures visible in AIA movies \citep{Li2016,Su2014}. With spectrographs (such as the Extreme-ultraviolet Imaging Spectrometer (EIS - \citet{Culhane2007}) aboard Hinode, the Interface Region Imaging Spectrograph~\citep[IRIS]{DePontieu2014} or from ground-based observatories) such motions were observed as blueshift on one side, and redshift on the other side of vertical columns in prominences. This suggests twisted magnetic structures or tornadoes \citep{Orozco2012,Su2014,Levens2016,Yang2018}. 
 
The tornado model of \citet{Luna2015} attempted to create a scenario where the prominence plasma is supported in a twisted magnetic field. However, rotation of prominence columns reported in AIA movies could correspond to an incorrect interpretation. \citet{Schmiederz2017} determined the true trajectory of an apparent helical prominence by reconstructing the velocity vectors of plasma blobs along the helical structure with IRIS data. These vectors with Dopplershifts equal to 50~km~s$^{-1}$ or more and transverse flows of only 5~km~s$^{-1}$ indicated that the structure was not helical but consisted of an horizontal magnetic field parallel to the solar disc. The apparent helical structure was due to perspective effect.

Dopplershift patterns can also be misleading if the field of view does not cover the entire rotating structure or if the temporal resolution is not fine enough. Using IRIS spectral data, \cite{Yang2018} presented a pattern of blue and redshifted velocities along the slit of the instrument, suggesting rotation around the axis of the prominence. However, it is difficult to confirm this rotating motion due to the small signal to noise ratio of the observation.
\citet{Schmieder2017} derived Dopplershift maps from H$\alpha$ observations of a tornado-prominence observed with the Multi subtractive Double pass spectrograph (MSDP) \citep{Mein1991} and demonstrated that the Dopplershift pattern evolved rapidly even though large areas of the prominence displayed coherent constant velocities with blue and red shifts. They did not confirm the rotation of the structure. Tornado-like structures observed over prominences as they cross the limb remain enigmatic.

A simultaneous multi-instrumental observation campaign comprising of IRIS, Hinode, the MSDP spectrograph (operating at the Meudon solar tower (MST)), SDO/AIA and other observatories focussed on a solar prominence which had manifested on the south-west (S51) solar limb on the 19th April 2018. This prominence corresponds to one anchorage-footpoint of a long east-west filament visible in H$\alpha$ on 16 April, with dark equidistant bushes along its axis (data in BASS2000.com). 

This prominence is very active with a tornado-like structure at its top. This joint observation provides a good opportunity to address several questions. Is the rotating motion real or is it an apparent motion? How is the global magnetic configuration of a tornado?
What is the nature of the plasma in tornado-like prominences?

In this paper, we present the data used in this study (Sect.~\ref{sec:instruments}); the large-scale evolution of the prominence in multiple temperatures (Sect.~\ref{sec:evolution}); with a focus on the dynamics of the tornado-like prominence and its global magnetic configuration (Sect.~\ref{sec:tornado}). Then, we present the analysis of its plasma parameters using H$\alpha$ and \ion{Mg}{ii} lines and comparison with 1D radiative transfer models (Sect.~\ref{sec:plasma}). In Sect.~\ref{sec:conclusion}, we discuss the results of the flows in the frame of magnetic support configuration. The high electron density of the plasma found at the top of the prominence could correspond to a dynamical phase of formation of the tornado-line prominence.

\section{Instruments}
\label{sec:instruments}
\subsection{IRIS}
The Interface Region Imaging Spectrograph~\citep[IRIS]{DePontieu2014} is a space-based multi-channel imaging-spectrograph.
IRIS provides observations in two far ultra violet channels (FUV, 1332-1358\AA~and 1390-1406\AA) and a near ultra violet channel (NUV, 2785-2835\AA).
These channels include strong chromospheric (\ion{Mg}{ii}, C II) and transition region (Si IV) lines.
Context pertaining to the position and surroundings of the slit can be found via observations from the Slit Jaw Imager (SJI) with three filters centred on 1330\AA, 1400\AA, and 2796~\AA\ respectively. The FUV filters have a bandpass of 54\AA, and the NUV 4\AA.

\begin{figure*}[ht!]
\centering
\includegraphics[scale=0.70]{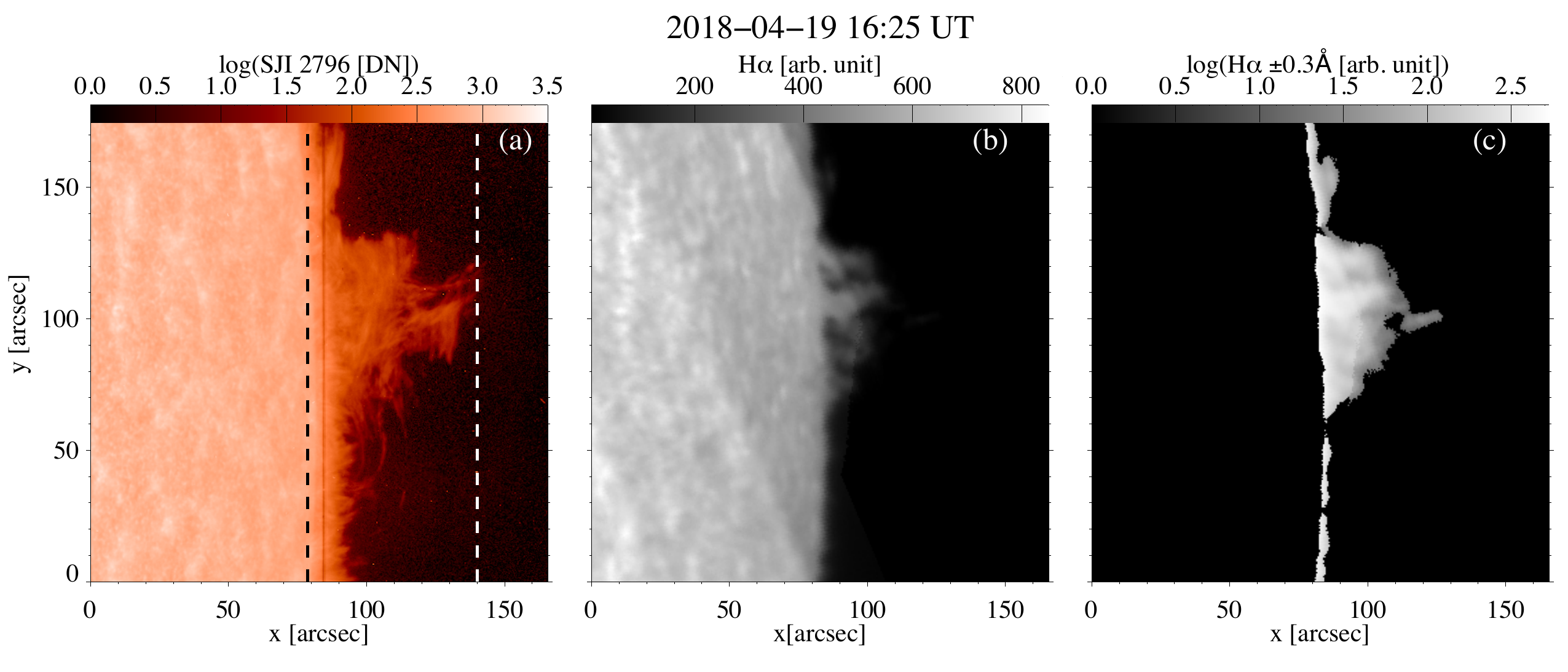}
\caption{Prominence observed by IRIS and the MSDP at the south-west limb of the Sun on 2018 April 19. The panels show: (a)~slit-jaw image of the \ion{Mg}{ii} line (SJI 2796), (b)~H$\alpha$ intensity in the line centre observed by MSDP showing the solar disk  used to co-align MSDP and IRIS maps}, 
(c)~H$\alpha \pm 0.3$~\AA\ intensity observed by MSDP. The dashed lines in the panel (a) marks the field of view of the IRIS raster.\label{figs:sji_msdp}
\end{figure*}

We focus on the simultaneous \ion{Mg}{ii} raster and SJI 2796~\AA\ observations of the prominence obtained on 19 April 2018 between 14:13 UT and 19:15 UT.
We mainly use the raster data of the \ion{Mg}{ii} 
lines to provide the plasma diagnostics (intensity, Doppler velocity, and FWHM).
We did not use the \ion{C}{ii} and \ion{Si}{iv} data because of the artifact of the aperture of the telescope which masks partly the field of view \citep{Wulser2018}. Eighteen very large coarse 32-step rasters were recorded during the five hours of observations. It took 16 minutes to perform one raster scan. During this time, eight SJIs were obtained (see an example in Fig.~\ref{figs:sji_msdp}a). The SJIs 
allowed us to study the fast transverse dynamics of the prominence material.

The details of the IRIS observations are summarised in Table~\ref{tab:instr_overview}. The data were downloaded from the IRIS database\footnote{\url{https://iris.lmsal.com/search/}.}. We used IRIS level-2 data corrected for the dark current, flat field and geometric distortion \citep{DePontieu2014}.

IRIS had a clockwise satellite rotation angle of 51 degrees, such that the solar limb was parallel to the y-axis of the instrument. We adopt this rotation angle for the co-alignment with the other instruments (AIA and MSDP). We note that it is easier to rotate AIA images than IRIS spectra data to minimise the influence of the rotation for the plasma diagnostics.

\subsection{MSDP}
To study the H$\alpha$ line, we use the observations obtained with the MSDP spectrograph operating in  the Meudon solar tower  at the  Paris Observatory.
The observations were obtained between 12:05 UT and 16:35 UT with sequences lasting 15 minutes each and an exposure time of 160~ms.
The observations consisted of individual bands, each of them covers a field of view of about 370 arcsec $\times$ 60 arcsec with a pixel size approximately equal to 0.5 arcsec.
Five adjacent bands recorded in 30s allowed to recover an image of the full field of view (370 arcsec $\times$ 270 arcsec) after processing the data with the MSDP software \citep{Mein1977,Mein1991,Mein2001}. Our prominence was only covered by two individual bands, one covering the main part of the prominence and the other band the top.
Each band is reconstructed from elementary spectral images of the entrance window (open slit) of the spectrograph obtained along a wavelength range ~$\pm 0.7$\AA.
The scattered light was reduced by 
subtracting line profiles recorded outside the Sun in the close vicinity of the prominence. This method takes advantage of the fact that wavelengths are almost constant along lines perpendicular to the dispersion.

Our work focuses on the last sequence between 16:21 UT and 16:35 UT which corresponds to an active phase of the prominence.
The intensity in H$\alpha$ line centre allows us to see the whole prominence and the bright pattern of the chromosphere which is very useful for co-alignment with the other instruments (Fig.~\ref{figs:sji_msdp}b). In this panel, we did not correct the mean brightening between the bands to show the discontinuity between them. This highlights the rotation angle of $\sim 30$~degrees between the direction of the band and the limb. For the further analysis, we used the H$\alpha$ maps with corrected intensity. Intensity and Dopplershift in each pixel in the prominence were computed in the range H$\alpha \pm 0.3$ \AA\ (Fig.~\ref{figs:sji_msdp}c and Fig.~\ref{figs:figs:msdp}b). We will discuss in detail how we proceed for the co-alignment of MSDP and IRIS observations. Table~\ref{tab:instr_overview} summarises the details of the MSDP data. The MSDP Meudon observations of the prominence in H$\alpha$ line are stored in the archive (LESIA08) of the Paris Observatory in Meudon.

\subsection{SDO/AIA}
To study the spatio-temporal dynamics of the prominence with high-resolution, we use data obtained by the space-based mission Solar Dynamics Observatory \citep[SDO;][]{Pesnell2012}.
The Atmospheric Imaging Assembly~\citep[SDO/AIA;][]{Lemen2012} instrument on-board SDO provides full-disc images that cover the solar atmosphere from the photosphere to the corona.
The seven EUV channels (e.g. AIA 304 \AA, AIA 171 \AA, AIA 193 \AA) provide observations with a nominal spatial resolution of 1.2 arcsec. (pixel size 0.6 arcsec) and a temporal scale of 12 sec.
In our analysis, we use the AIA data obtained in the same time interval as the IRIS data.

We use AIA 304 \AA\ filter where the main emission comes from the \ion{He}{ii} line at 303.78 \AA\ formed by scattering of photons on ionised helium at a temperature of $\log T=4.7$ \citep{Labrosse2010}. 
Furthermore, there is a number of lines formed at coronal temperatures in the bandpass of 304 \AA\ \citep{Dere1997,Landi2012}. However, they weakly affect the emission of cool prominences.

We also studied AIA 171 \AA\ and AIA 193 \AA\ images, formed at temperatures of $\log T=5.8$ and $\log T=6.1$ respectively. 
In these wavelengths, the cool plasma in the corona absorbs the background coronal emission and is visible in absorption. This was discovered by observations of the fine structures of a filament with the SST telescope \citep{Scharmer2003} when compared with a TRACE image at 171 \AA\ \citep{Schmieder2004}. 
 Prominences are well observed in the two filters of AIA 171 \AA\ and 193 \AA\ as absorption structures. In AIA 304 \AA, prominences show a completely difference appearance, because they are seen in emission.

\begin{table*}[ht!]
\caption{Details of IRIS, MSDP and SDO/AIA observations made on 19 April 2018.}\label{tab:instr_overview}
\resizebox{\textwidth}{!}{\begin{tabular}{@{}lcllll@{}}
\hline
Instrument & \multicolumn{1}{l}{Observation time} & Spectroscopic measurements & & Imaging & \\ \hline
\multirow{8}{*}{IRIS} & \multirow{8}{*}{\begin{tabular}[c]{@{}c@{}}14:13-19:15 \\ UT
\end{tabular}} & Pointing: & 633"; -753" & & \\
 & & Field-of-view (FOV): & 62"$\times$175" & FOV: & 167"$\times$175" \\
 & & Observation repetition: & 18 & Image numbers & 144 \\
 & & Steps: & 32 & & \\
 & & Step cadence: & 31.4 s & & \\
 & & Raster cadence: & 1.005 s & & \\
 & &  Spatial pixel size: & 0.33" &  Spatial pixel size: & 0.167" \\
 & & Line & \ion{Mg}{ii} k, \ion{Mg}{ii} h & & \ion{Mg}{ii} (SJI 2796) \\ \hline
\multirow{5}{*}{MSDP} & \multirow{5}{*}{\begin{tabular}[c]{@{}c@{}}12:03-16:35 \\ UT
\end{tabular}} & FOV: & 370"$\times$270" & & \\
 & & Observation repetition: & 300 & & \\
 & & Time resolution: & 30 sec & & \\
 & & Spatial pixel size: &  0.5" & & \\
 & & Line & H$\alpha$ & & \\ \hline
\multirow{4}{*}{SDO/AIA} & \multirow{4}{*}{\begin{tabular}[c]{@{}c@{}}14:13-19:15 \\ UT
\end{tabular}} & & & FOV & 600"$\times$600" \\
 & & & & Time resolution & 12 sec \\
 & & & & Spatial pixel size &  0.6" \\
 & & & & Line & HeII (AIA 304 \AA) \\ \hline
\end{tabular}}
\end{table*}

\begin{table*}[ht!]
\centering
\caption{Properties of observed spectral lines.} \label{tab:line_properties}
\begin{tabular}{@{}lllll@{}}
\hline
Instrument & Wavelength [\AA] & Line/Band & log $T$ {[}K{]} & Atmospheric regime \\\hline
 & 2796 & \ion{Mg}{ii} k & 3.6 - 3.9 & Chromosphere \\
IRIS & 2803 & \ion{Mg}{ii} h & 3.6 - 3.9 & Chromosphere \\
 & & \ion{Mg}{ii} k -peak & 3.9 & Upper chromosphere \\ \hline
MSDP & & H$\alpha$ & 3.9 & Upper chromosphere \\ \hline
 & 304 & \ion{He}{ii} & 4.7 & Upper chromosphere, TR \\
SDO/AIA & 171 & \ion{Fe}{ix} & 5.9 & Upper TR \\
 & 193 & \ion{Fe}{xii}, \ion{Fe}{xxiv} & 6.1 - 7.3 & Corona \\\hline 
\end{tabular}
\end{table*}

We use pre-processed SDO/AIA data, provided by the Joint Science Operations Center (JSOC\footnote{\url{http://jsoc.stanford.edu}.} that correspond to level-1.5.
The SDO/AIA data exported from JSOC were mutually co-aligned with a high spatial accuracy.

\section{Evolution of the large-scale prominence in different temperatures}
\label{sec:evolution}

The MSDP, IRIS, and AIA instruments provide a view of the prominence in multiple temperatures. Due to opacity and different ionisation temperature effects the prominence shape observed in each wavelength is different, revealing different structures.

\subsection{H$\alpha$, \ion{Mg}{ii} and AIA 304 prominence}
\begin{figure*}[ht!]
\centering
\includegraphics{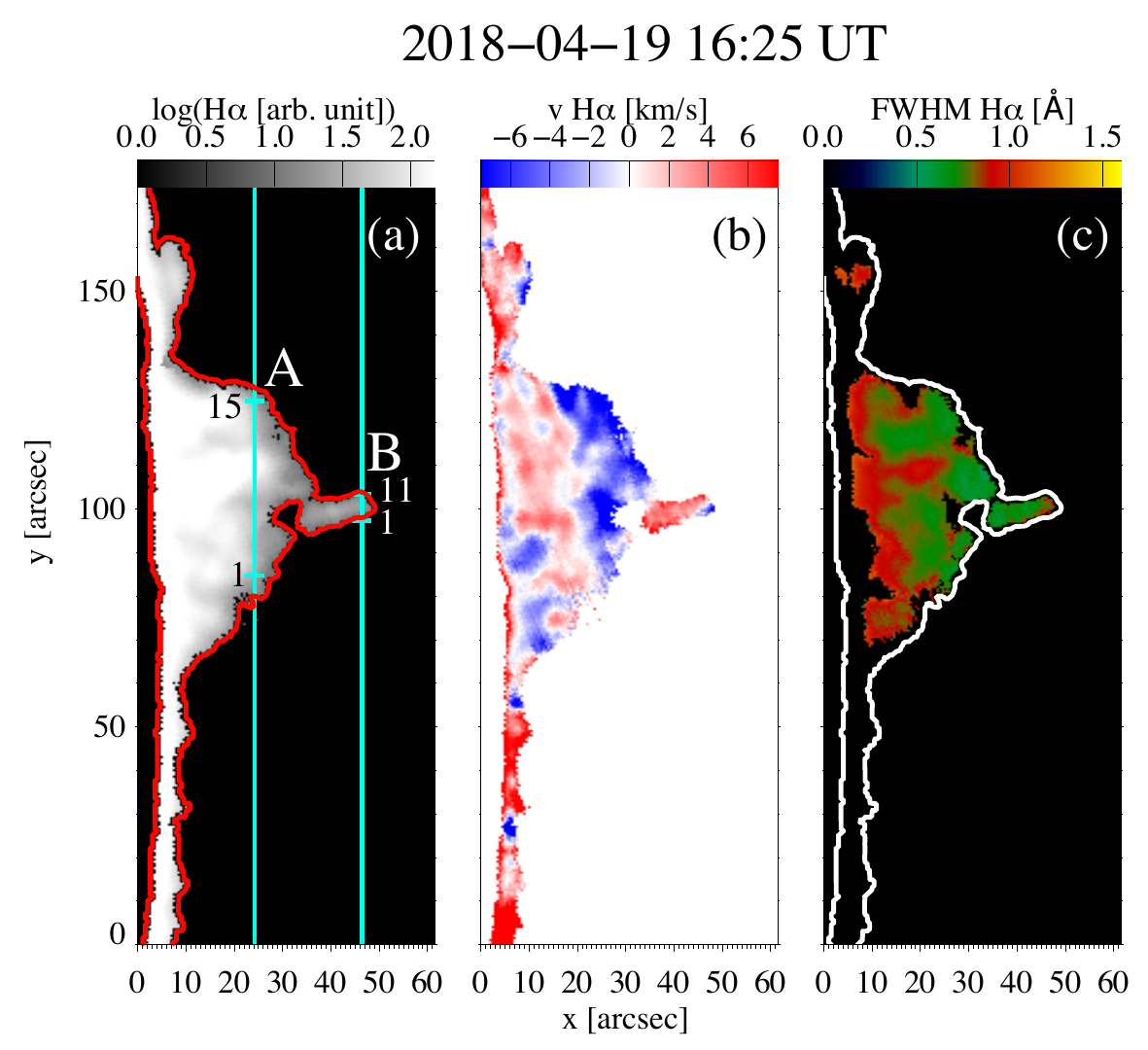}
\caption{Prominence diagnostic based on the MSDP observation. (a) H$\alpha$ intensity at the line centre, (b) Doppler velocity, (c) FWHM. The contours present the H$\alpha$ prominence defined as Mask-H$\alpha$ (see Sect.~\ref{sec:mask}).
The blue vertical lines, in panel (a), are the two selected slit positions (A, B) used for analysing the profiles (Fig.~\ref{fig:MSDP_profiles1}, Fig.~\ref{fig:MSDP_profiles2}) and used for plasma parameter diagnostic (Table~\ref{tab:mg2_ha_comparison}). The small horizontal lines are the positions corresponding to the profiles A1-A15 and  B1-B11.
}
\label{figs:figs:msdp}
\end{figure*}

The global shape of the prominence observed in H$\alpha$ and in IRIS SJI 2796~\AA\ is very different (Fig.~\ref{figs:sji_msdp}a and Fig.~\ref{figs:figs:msdp}a). In H$\alpha$, the prominence consists mainly of two narrow and low columns with weak emission loops joining the column to the solar surface (Fig.~\ref{figs:sji_msdp}c, Fig.~\ref{figs:figs:msdp}a) while the IRIS prominence has a wide base of 100 arcsec. parallel to the y-axis and a height of about 50 arcsec. (37500~km). The top of IRIS prominence is narrower with one or two horns depending on the time. A prominence with such an appearance was discussed in \citet{Wang2016}. The differences in apparent morphology of prominences observed in different chromospheric lines have been discussed in the past. The optical thickness of the \ion{Mg}{ii} h or k line is $\sim$100 times greater than that of H$\alpha$ \citep{Ruan2019}. This explains why all the low density structures can be better seen in \ion{Mg}{ii} \citep{Heinzel2015,Levens2016}. Due to contrast issues, it is not possible to visualize the low emission structures in H$\alpha$.

The prominence in H$\alpha$ is defined with 'Mask-H$\alpha$' (see Sect.~\ref{sec:mask}).
Dopplershift and FWHM can be computed in all the area limited by this mask. The area of the computed Dopplershifts corresponds to a large part of the IRIS prominence (Fig.~\ref{figs:sji_msdp} panel a, Fig.~\ref{figs:figs:msdp} panel b).

\begin{figure*}[ht!]
\centering
\includegraphics{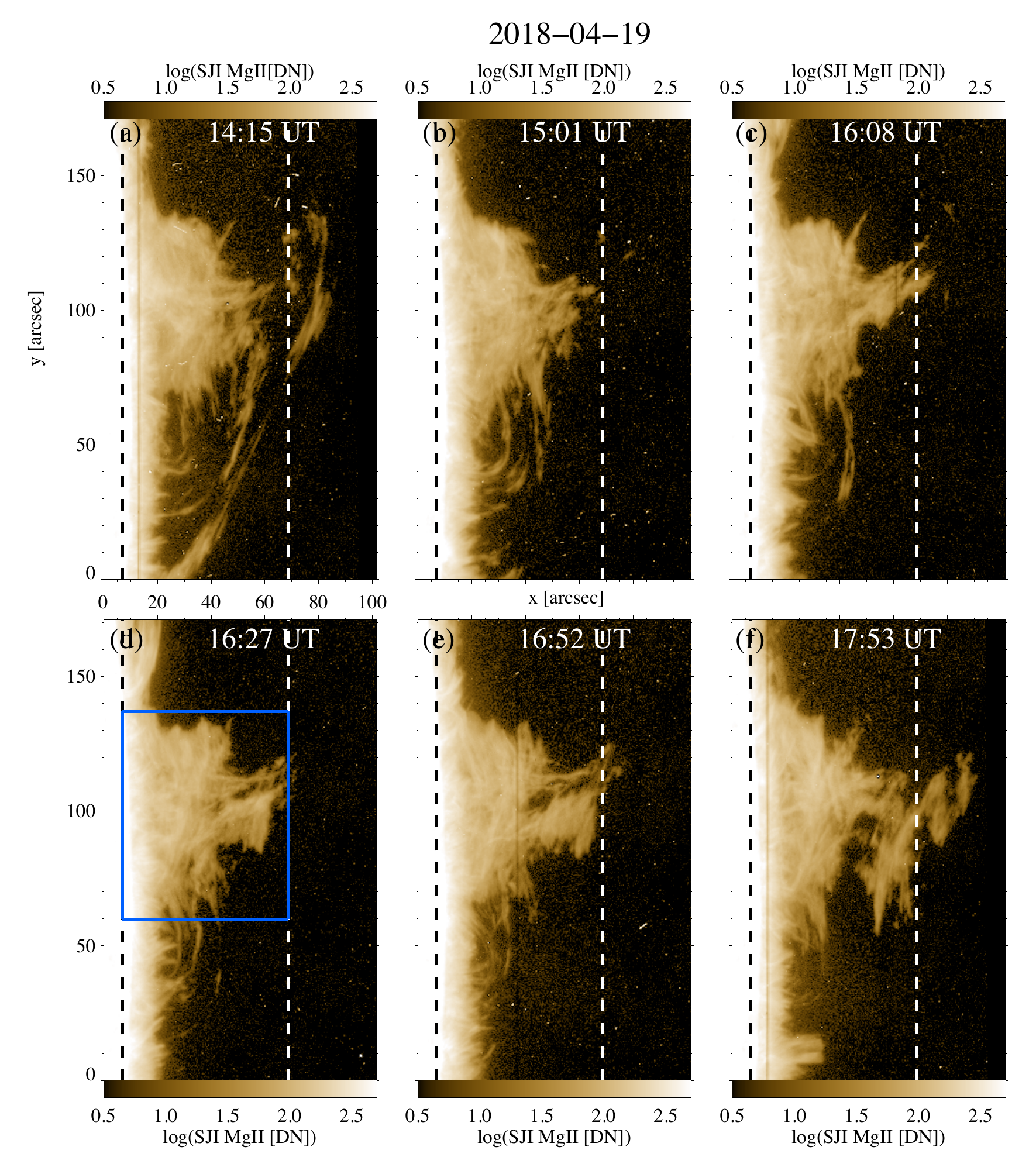}
\caption{Spatio-temporal evolution of the prominence intensity in \ion{Mg}{ii} slit-jaw 2796 \AA\ images obtained by IRIS between 14:15 UT to 17:53 UT. The dashed vertical lines mark the IRIS raster field of view.
The blue box in panel (d) define  the FOV used for detailed study of the velocity distribution (Fig.~\ref{figs:his_dop_vel}).}
The temporal evolution is available as an online movie (Movie1).
\label{figs:overview_sji}
\end{figure*}

The IRIS SJI 2796~\AA\ movie (Movie1) shows tremendous moving features during the five hours of observing time. During this time the global structure of the prominence rises slowly ($<3$~km~s$^{-1}$). The different periods of activity suggest the need for deeper analysis. In the IRIS SJI movie, we see major changes at the top of the prominence. The evolution of the prominence is shown in six panels in Fig.~\ref{figs:overview_sji}. At the beginning of the observation (14:15 UT) we observe long narrow loops joining the top of the prominence to the solar surface.

Between 15:45 and 17:30 UT the top appears as a relatively narrow column which displays some twisted motions. After turning to the North and South, around 17:30 UT material from the top detaches, untwists and unwinds to ultimately expand in a horizontal direction. Then, between 17:56 to 18:18 UT the material falls following long loops towards the southern limb.
 
We estimated the transverse flows along these loops to be less than 7 km~s$^{-1}$.

\begin{figure*}[ht!]
\centering
\includegraphics{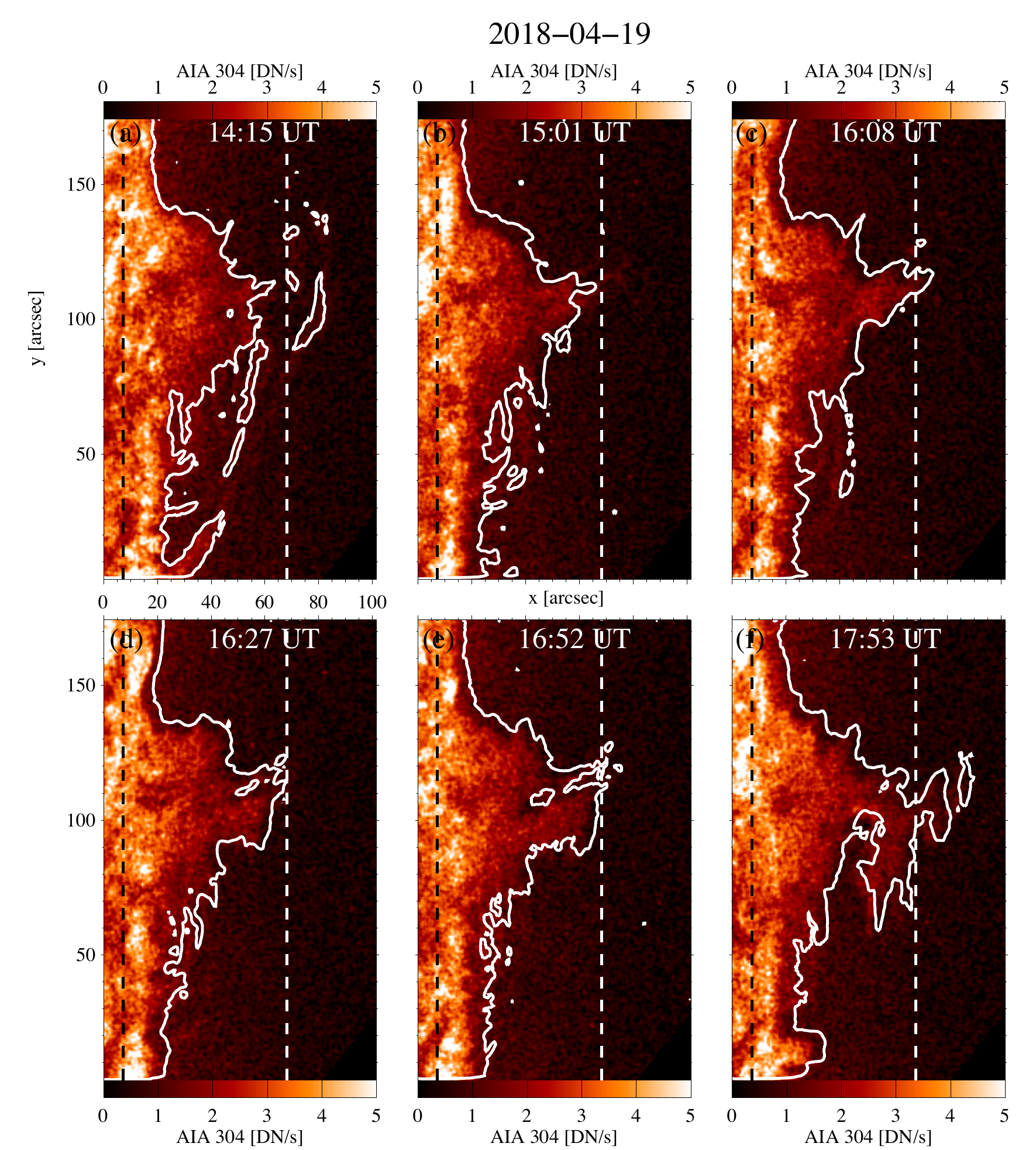}
\caption{Spatio-temporal evolution of the AIA 304 \AA\ intensity observed between 14:15 UT to 17:53 UT. The vertical dashed lines mark the IRIS raster field of view. The overlaid contour corresponds to the contour of the IRIS SJI 2796, level=1. 
The temporal evolution is available as an online movie (Movie2).
\label{figs:overview_aia304}}
\end{figure*}

The dynamics of the prominence in AIA 304 movie (Movie2) has a similar behaviour to the prominence in the IRIS SJI 2796~\AA\ movie but with less contrast (Fig.~\ref{figs:overview_aia304}). The fine structures, particularly at the top of the prominence, are not well observed. Their morphology changes rapidly and can only be resolved with the high spatial resolution of IRIS.

The AIA 304 filter contains the \ion{He}{ii} 303.78 \AA\ line which is strongly affected by opacity effects, similar to the \ion{Mg}{ii} lines. Prominences observed in this line look completely different from how they are seen in H$\alpha$ but similar to what is seen in IRIS SJI 2796. 
 The optical thickness of \ion{He}{ii} 303.78~\AA\ is between 10$^2$ and 10$^3$ \citep{Levens2016}. This implies that we mainly observe the structures located at the front of the prominence. This allows us to observe weakly emitting structures not visible in H$\alpha$ \citep{Ruan2018} resulting in a more extended appearance of the prominence similar to what is seen in the IRIS SJI.
 
\subsection{AIA prominence in 171 \AA\ and 193 \AA}
\begin{figure*}[ht!]
\centering
\includegraphics{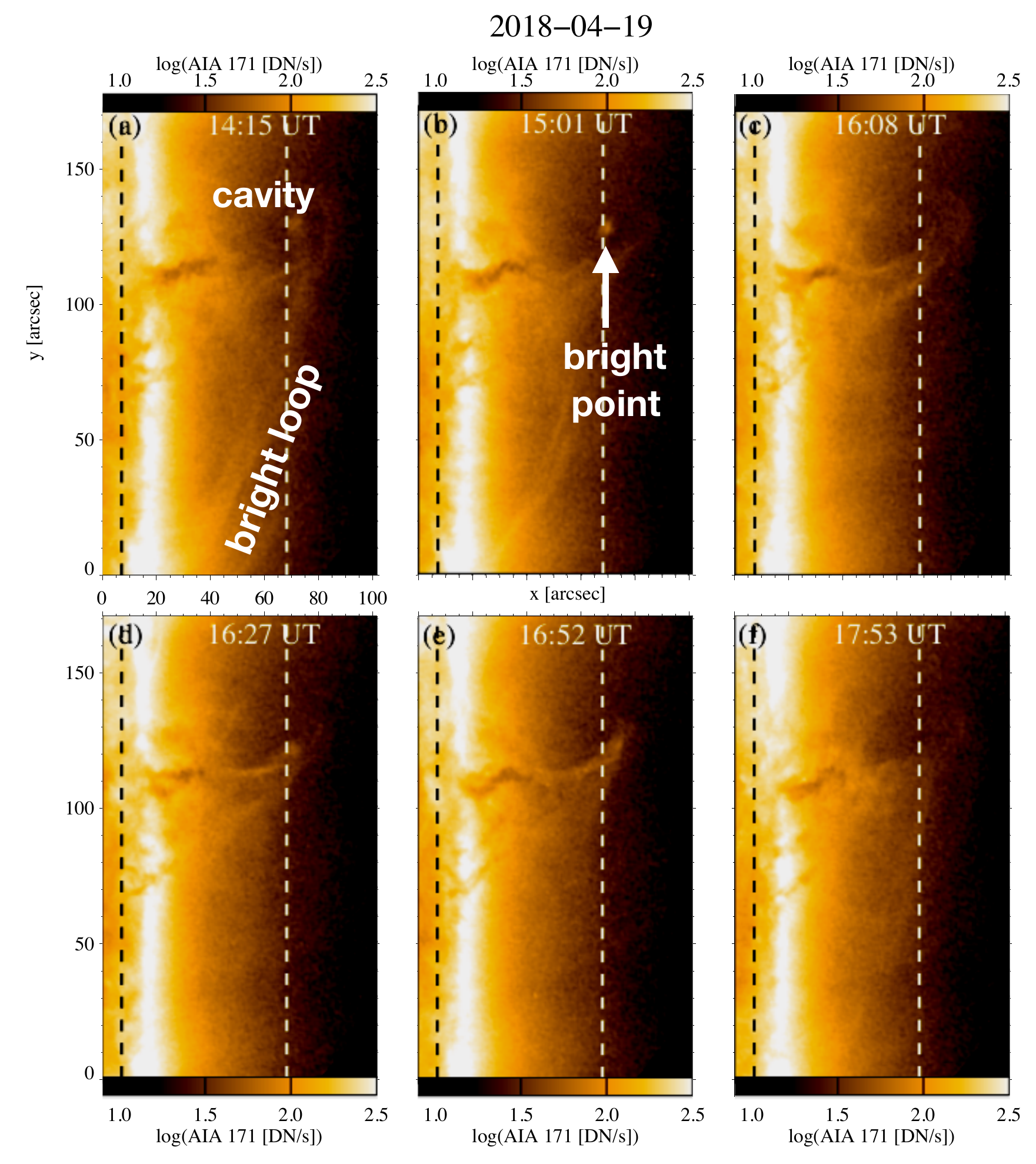}
\caption{Spatio-temporal evolution of the AIA 171 \AA\ intensity observed between 14:15 UT to 17:53 UT.
 The vertical dashed lines mark the IRIS raster field of view. We note the presence of the cavity and long extended bright loop in (a), a bright point in (b). The cavity is visible in all the panels and the dark twisted column is surrounded with emission. The temporal evolution is available as an online movie (Movie3).
\label{figs:overview_aia171}}
\end{figure*}

\begin{figure*}[ht!]
\centering
\includegraphics{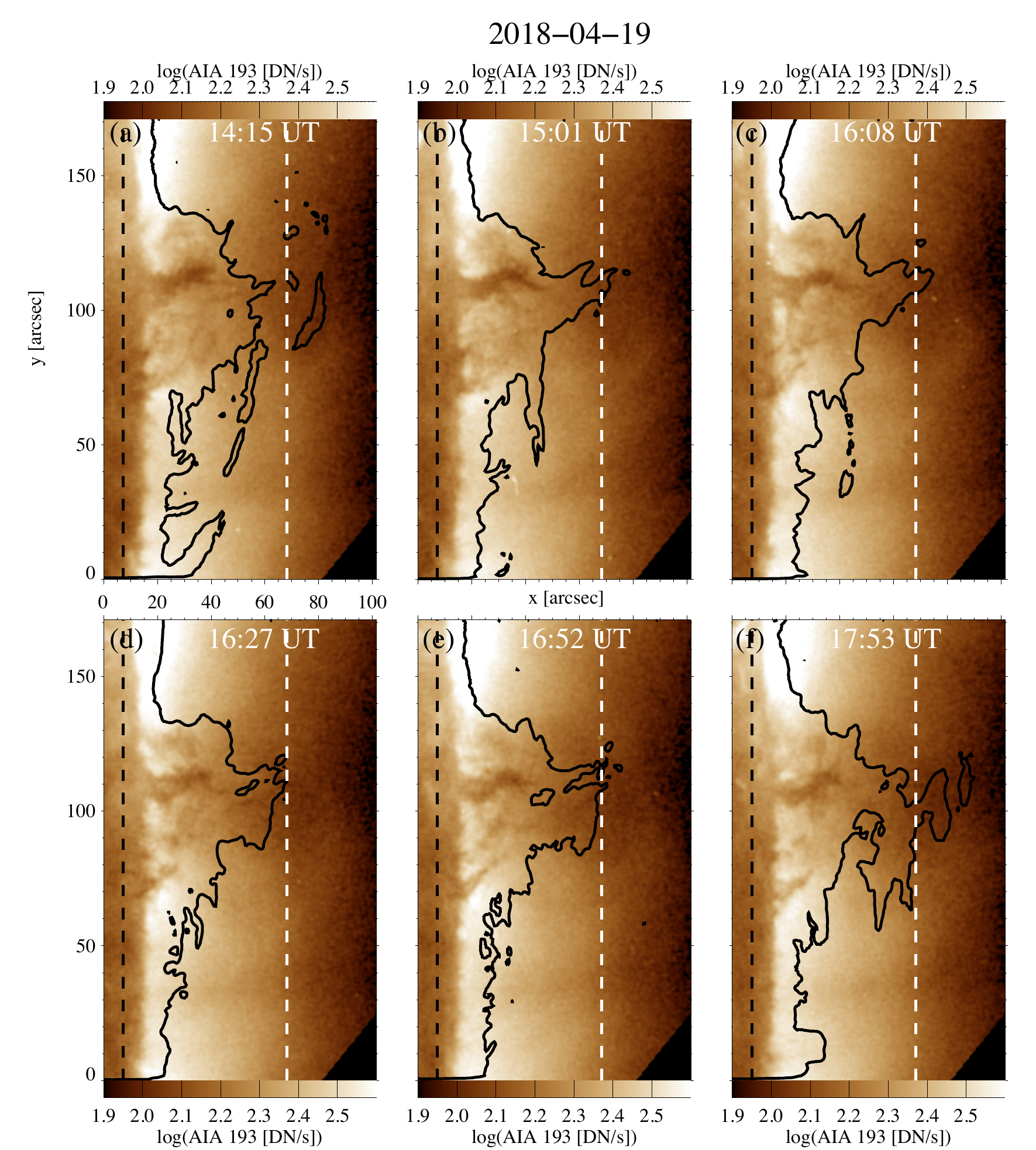}
\caption{Spatio-temporal evolution of the AIA 193 \AA\ intensity observed between 14:15 UT to 17:53 UT.
 The vertical dashed lines mark the IRIS raster field of view. The overlaid contour corresponds to the contour of the IRIS SJI 2796, level=1. The dark features (columns and fuzzy dark material) observed in the prominence correspond to the prominence observed in H$\alpha$. The temporal evolution is available as an online movie (Movie4).
\label{figs:overview_aia193}}
\end{figure*}

In the SDO/AIA 171 \AA\ and 193 \AA\ images, the prominence absorbs the background coronal emission and appears as a dark column located in the centre of the prominence, oriented perpendicularly to the solar limb (Figs.~\ref{figs:overview_aia171} and \ref{figs:overview_aia193}). This absorption structure corresponds to the similarly shaped brightest region in the H$\alpha$ MSDP image in Fig.~\ref{figs:sji_msdp} (b, c). The signal in AIA 171 is dominated by the \ion{Fe}{ix} 171 \AA\ line formed typically at $\log T =5.9$. However, this line is also sensitive to plasma at a lower temperature of $\log T < 5.6$ in the prominence-to-corona transition region \citep{Parenti2012}. This is responsible for the diffuse emission surrounding the central absorption column, and the illumination of the extended loops joining the limb in the south and the loops surrounding the cavity. The SDO/AIA 193 channel emission is multi-thermal ($\log T=6.1-7.3$) and consists of contributions of several coronal lines of \ion{Fe}{xii} and \ion{Fe}{xxiv}. In this channel, the prominence is visible purely in absorption from the hydrogen and helium continuum opacity. The optical thickness of the continua at these wavelengths is comparable to the optical thickness of the H$\alpha$ line which is around unity \citep{Schmieder2004,Anzer2005}. In the area surrounding the dark column, smaller columns are visible (Fig.~\ref{figs:overview_aia171}). These wispy structures are less extended than in AIA 304 and are located inside the contour of the IRIS prominence. In the AIA 193 movie (Movie4), the top of the dark absorption column is seen to oscillate and change shape giving the impression of some twist.

In the AIA 171 movie (Movie3), a part of a loop is seen surrounding the cavity (Fig.~\ref{figs:overview_aia193}). At 15:01 UT a bright point is seen in the middle of the cavity. Is this bright point due to reconnection leading to activity in the prominence? It is not directly related to accelerated motions, and no emission is detected in the X-ray telescope (XRT) images onboard the Hinode solar telescope \citep{Golub2007}. Later, more extended loops join the main body of the prominence to the solar surface. The top of the prominence becomes elongated in an orientation parallel to the limb, with material flowing towards the north and south. This gives also the impression of rotating structures. 

\section{Tornado-prominence dynamics}\label{sec:tornado}

\begin{figure*}[ht!]
\centering
\includegraphics[scale=0.75]{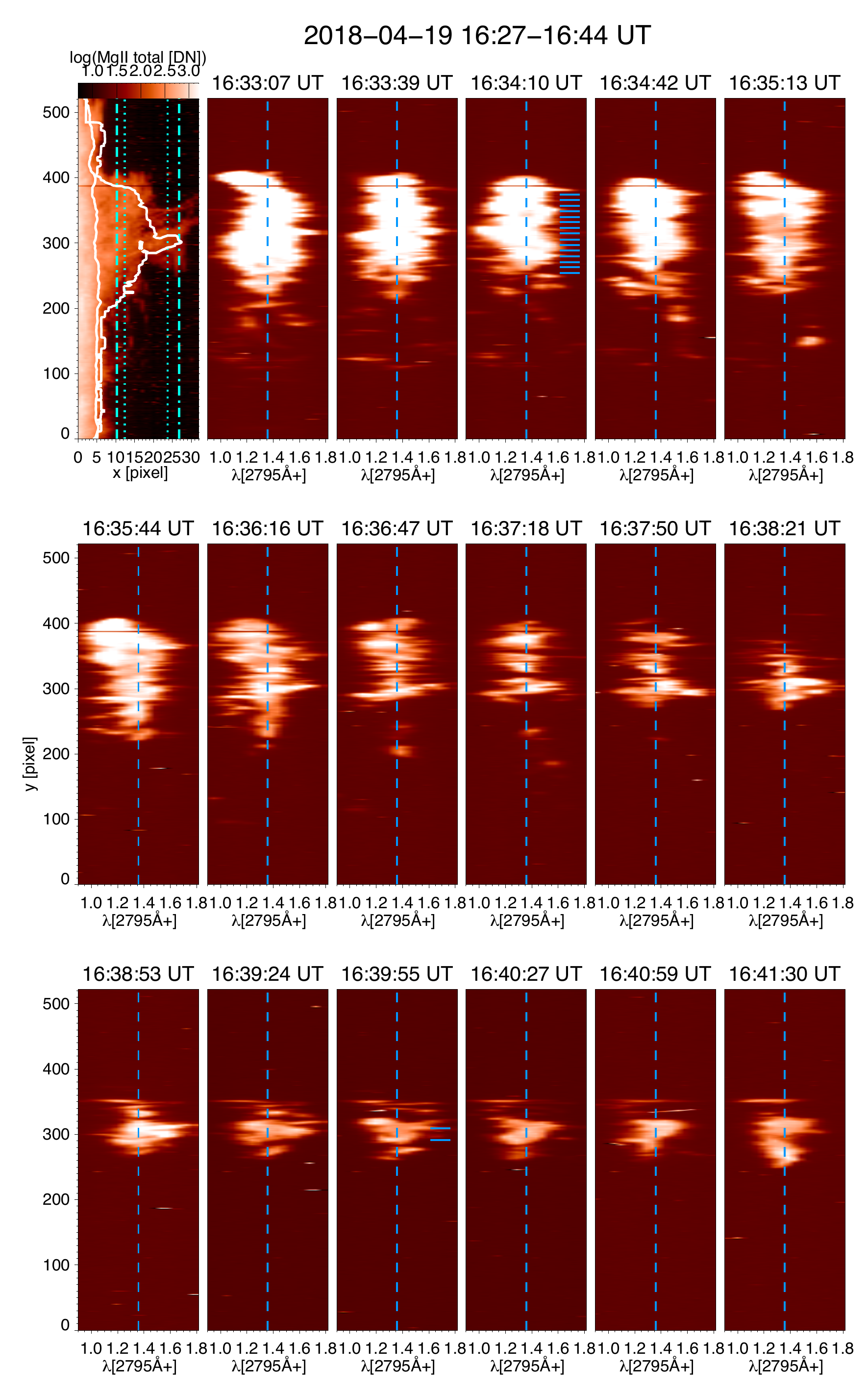}
\caption{Prominence spectra of raster 8. In the first row, the left panel presents the reconstructed map using the integrated intensity of \ion{Mg}{ii} k line, the five following panels are the spectra along the slit in positions 10 to 14, the second row in positions 15 to 20, in the bottom row in positions 21 to 26. In the upper left panel, the dotted dashed vertical lines represent the extreme slit positions (10 and 26) for which the spectra are shown in this figure. The dotted vertical lines are the two selected slit positions used for analysing the 
profiles and used for plasma parameter 
diagnostic. In panels of the spectra, the vertical dashed blue line corresponds
to the rest velocity. 
In the panel at 16:34:10 UT (slit position 12) the horizontal lines are the positions corresponding to the profiles A1-A15 drawn in Fig.\ref{figs:spectrum1} (Appendix~\ref{sec:appendix}). In panel at 16:39:55 UT (slit position 23) the horizontal lines limit the domain where the profiles (B1-B11) are shown in Fig.~\ref{figs:spectrum2} (Appendix~\ref{sec:appendix}).
The unit of the y-axis is in pixel (0.33 arcsec).
\label{figs:mg2_spectra}}
\end{figure*}

From the eighteen IRIS rasters obtained between 14:13 UT and 19:15 UT, we analysed the 3D dynamics of the activity in the prominence with a particular focus on the top which appears similar to a tornado at different times.
The eighteen rasters include 32 spectra of \ion{Mg}{ii} h and k (Table~\ref{tab:instr_overview}). Of these spectra, 28 cover the prominence, while the first four spectra are on the disk or at the limb including spicules.
Figure~\ref{figs:mg2_spectra} shows an example of the \ion{Mg}{ii} k spectra through the prominence with a reconstructed map obtained using the integrated intensity of \ion{Mg}{ii} k.

\subsection{\ion{Mg}{ii} : Wavelength calibration method}\label{sec:rest_velocity}

To analyse the IRIS data, we first calibrated in wavelength the \ion{Mg}{ii} spectra.
We assumed that the average velocity of the photospheric line \ion{Ni}{i} 2799.474~\AA\ is equal to zero.
A mean spectral profile of \ion{Ni}{i} was created for the whole slit position located on the solar disk (first slit position).
Using a spline interpolation, we computed the \ion{Ni}{i} peak position.

We confirmed this calibration by computing an average spectral profile of the \ion{Mg}{ii} k line over the disk.
The wavelength value of dip of the mean spectral \ion{Mg}{ii} k profile is consistent with the nominal rest wavelength of the \ion{Mg}{ii} k 
 line at 2796.35 \AA\ \citep{Pereira2013} and confirmed the \ion{Ni}{i} calibration result.

\subsection{\ion{Mg}{ii}: Gaussian fitting method}\label{sec:gauss_fit}

The observed \ion{Mg}{ii} spectra generally have no central reversal. Therefore a Gaussian fitting procedure may be applied to retrieve the total intensity, full-width at half-maximum (FWHM), and Dopplershift.
To avoid noise, we fitted with a single Gaussian profile, \ion{Mg}{ii} k profiles when the peak amplitude was three times larger than the standard deviation of the background continuum intensity. 
The background continuum intensity was defined in the waveband of 5.09\AA\, centered at 2811.23\AA.

To calculate velocity (v) with the Doppler effect, we used the following formula:
\begin{equation}
\label{eqn:eq1}
v=c\frac{\lambda_{observed}-\lambda_{emitted}}{\lambda_{emitted}}
\end{equation}
where $c$-speed of light, $\lambda_{observed}$-wavelength observed by IRIS, $\lambda_{emitted}$-wavelength emitted by plasma (theoretical wavelength).

\begin{figure*}[ht!]
\centering
\includegraphics{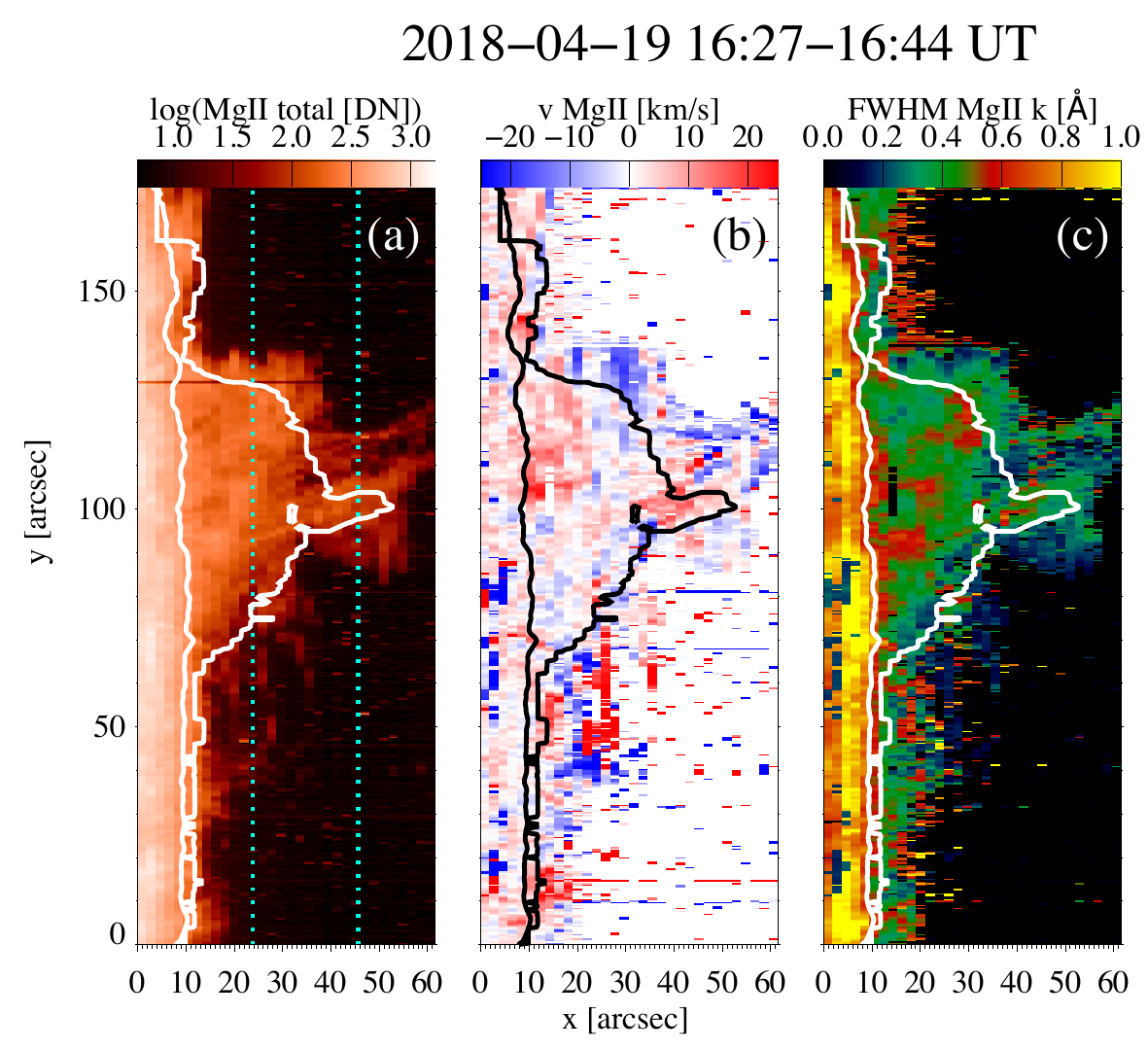}
\caption{Prominence reconstructed maps based on the IRIS spectra of \ion{Mg}{ii} k line (raster 8); (a)
integrated intensity of \ion{Mg}{ii} k line, (b) Doppler velocity, and (c) width (FWHM) of \ion{Mg}{ii} k line profile obtained by fitting the profiles with a single Gaussian function. The inner contour line (around $x=10$ arcsec) corresponds to the solar surface defined as the level 0.6~\AA~of the FWHM of \ion{Mg}{ii} k line; the outer contour corresponds to Mask-H$\alpha$ based on MSDP observation (see Mask-H$\alpha$
in Fig.~\ref{figs:figs:msdp}). 
This new contour defined by (Mask-H$\alpha$ - pixels on the disk)} will be used for the statistics of \ion{Mg}{ii} parameters in and out the H$\alpha$ prominence (see Sect.~\ref{sec:mask}).
The blue dotted lines, in the panel (a), are the two selected slit positions (A, B) used for analysing the profiles (Fig.~\ref{figs:spectrum1}, Fig.~\ref{figs:spectrum2} in Appendix~\ref{sec:appendix}) and used for plasma parameter diagnostic (Table~\ref{tab:mg2_ha_comparison}).
\label{figs:iris}
\end{figure*}

We used the fitting parameters to determine the position of the line centre and the corresponding total intensity, FWHM, and Doppler shift.
Figure~\ref{figs:iris} shows an example of these three quantities for raster 8.

\subsection{\ion{Mg}{ii}: Quantile method}
An alternative method to the Gaussian fitting is the quantile method \citep{Kerr2015, Ruan2018}.
We applied the quantile method only for raster points where the peak intensity of the \ion{Mg}{ii} k line was at least three times larger than the standard deviation of the background continuum. 
The quantile method involves calculating the cumulative distribution function of the intensity of individual line profiles  in a wavelength range.
Based on the cumulative histogram of intensity, we calculated the 0.12, 0.5 and 0.88 quantile (percentil) parameters (respectively Q1, Q2, Q3) considering a wavelength range of 2.04\AA\ centered on 2796.46\AA\ (this is the same wavelength interval as in the Gaussian fitting method). 
Based on the position of the line centre (peak), we created a map of peak intensity ($E_{\ion{Mg}{ii} k}^{Q; peak}$).
Again using the Eq.~\ref{eqn:eq1} as in Sect.~\ref{sec:gauss_fit}, we calculate the Doppler velocity.
Then, we calculated FWHM as the wavelength distance between 0.12 (Q1) and 0.88 (Q3) quantile.
The Gaussian and quantile methods give comparable results (Fig.~\ref{figs:his_dop_vel}).

\subsection{Evolution of the tornado-prominence viewed in IRIS raster}

\begin{figure*}[ht!]
\centering
\includegraphics{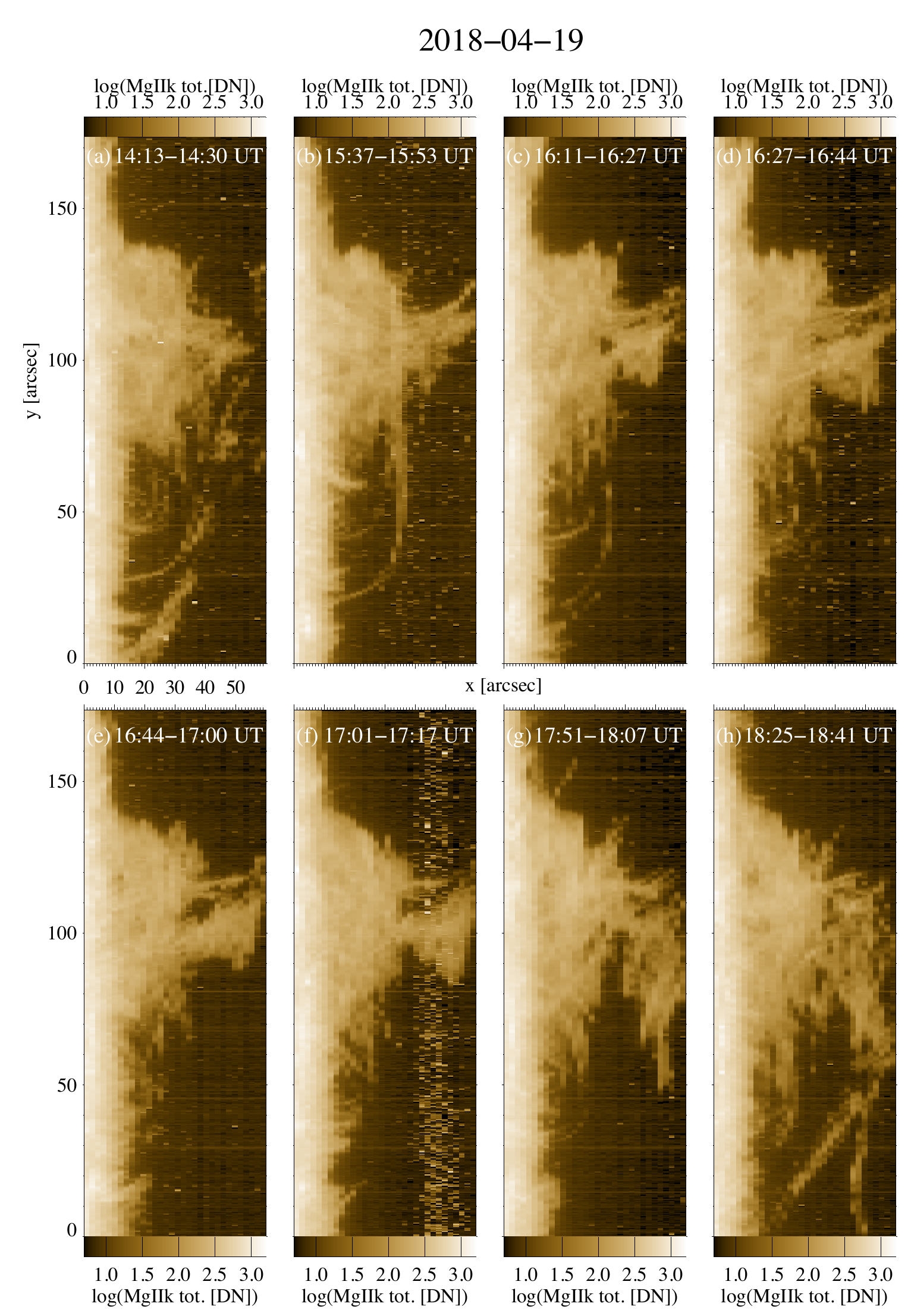}
\caption{Sample of integrated intensity maps in \ion{Mg}{ii} k line, each map being reconstructed from the 32 spectra of one raster, showing the evolution of the tornado at the top and the long loops connecting with the limb between 14:15 UT to 18:41 UT (rasters 1, 5, 7, 8, 9, 10, 13, 17).
\label{figs:iris_intensity}}
\end{figure*}

\begin{figure*}[ht!]
\centering
\includegraphics{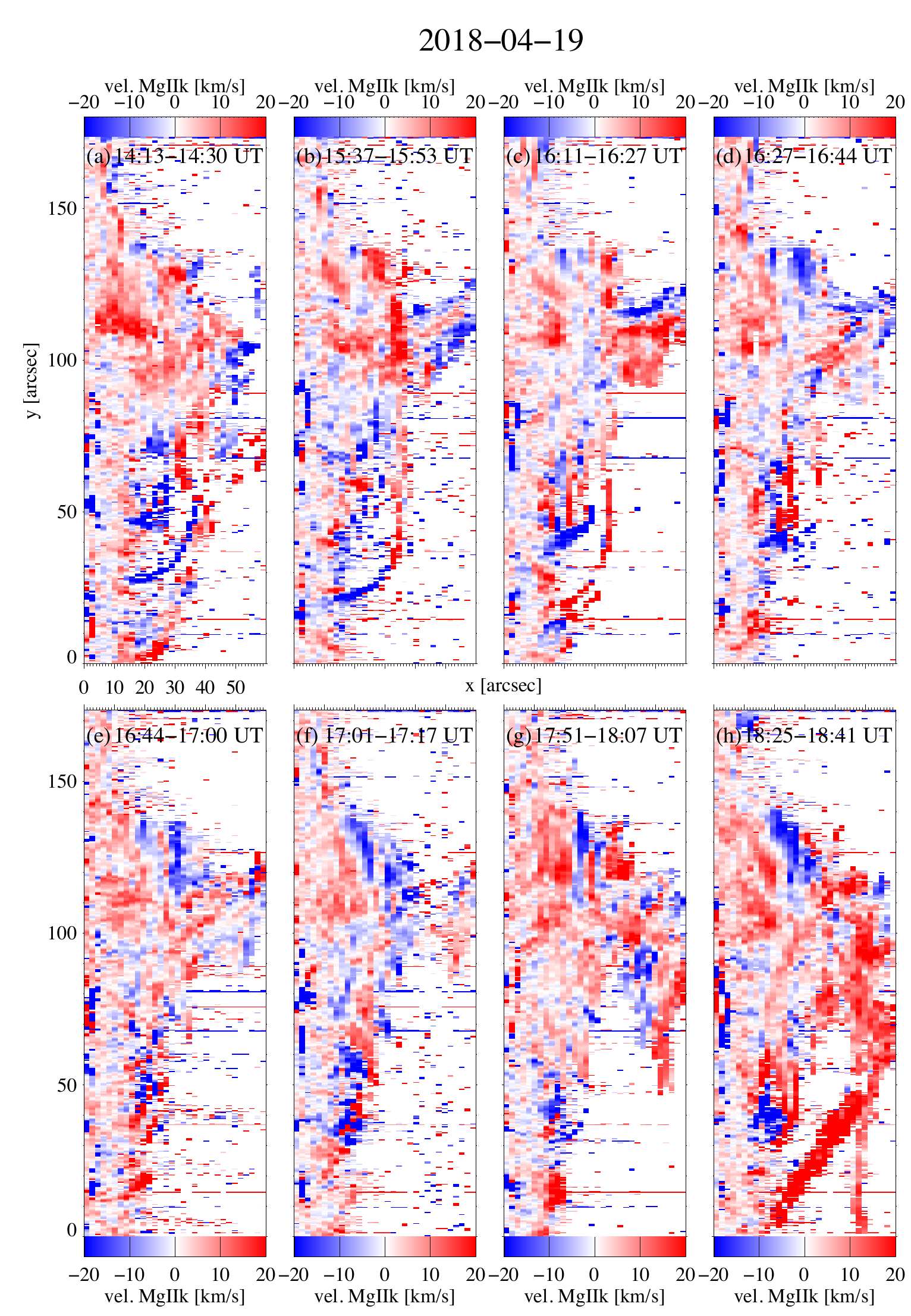}
\caption{Doppler velocity maps (Gaussian fitting) in \ion{Mg}{ii} k line computed from the spectra of the rasters observed by IRIS between 14:15 UT to 18:41 UT (rasters 1, 5, 7, 8, 9, 10, 13, 17). The temporal evolution is available as an online movie 
(Movie5).
\label{figs:iris_doppler}}
\end{figure*}

The movies (Movie1, Movie5) allow us to follow the evolution of the tornado in 3D with the associated Dopplershifts at a cadence of 16 min. This is a relatively low cadence for an active prominence.
Figure~\ref{figs:iris_intensity} and \ref{figs:iris_doppler} summarise the main behaviour of the tornado plasma
with intensity and simultaneous Dopplershift maps.

At the beginning of the observations around 14:13 UT we see long alternating blue and redshifted loops which join the top of the prominence to the disk. Between 15:27 and 16:27 UT the top of the prominence extends to higher altitudes and exhibits changes in Dopplershift with blue/red areas aligned perpendicularly to the limb. Around 16:07 UT, 16:35 UT to 16:52 UT a vertical pillar like a tornado is observed and is seen to rapidly develop. In the rasters 7, 8 and 9 (16:11 to 17:00 UT) redshifts and adjacent blueshifts are observed along the vertical pillar, simulating a tornado. At the end of raster 13 (around 17:53 UT), large flows with trajectories more or less parallel to the limb are observed from the top of the prominence ejecting blueshifted material towards the North and redshifted material towards the South. Finally, the material moving along long the loops, joining the top of the prominence to the south limb, mostly shows redshifts. This flow is found to be moving away from the observer (18:41 UT, raster 17).

\subsection{Co-alignment IRIS and MSDP data}
A two step procedure to co-align IRIS raster and MSDP data was undertaken.
First, we reduced the IRIS SJI 2796~\AA\ images resolution to the pixel size of MSDP (0.5 arcsec.).
Then, we used a cross-correlation method to co-align an IRIS SJI image with a MSDP intensity map of H$\alpha$ at line center.
The co-alignment was done with the solar disk using the structures of the network.
We obtained a precision of alignment better than 1 arcsec between IRIS SJI and MSDP image.

In the second step, we co-aligned IRIS SJI 2796~\AA\ images with IRIS raster data.
To achieve this, we used the IRIS raster integrated intensity maps in the \ion{Mg}{ii} k line and reduce the SJI resolution to the raster image.
Then, we used the fiducial as a spatial reference in the rasters and a cross-correlation method to align the SJI and raster image. 
A cross-correlation method was applied in the full field of view of the \ion{Mg}{ii} k raster intensity image. Due to fast evolution of the prominence we chose the last SJI image obtained during the rastering process which corresponds better to the prominence, the first one corresponds to the disk and the low part of the prominence.
The accuracy of the co-alignment is better than 2 arcsec. The result is shown in Fig.~\ref{figs:sji_msdp} (a, b, c)

\subsection{H$\alpha$ Dopplershifts}

In each pixel of the MSDP prominence field of view a H$\alpha$ profile is retrieved and Dopplershifts were computed with a bisector method. H$\alpha$ profiles in prominences are relatively narrow compared to profiles on the disk, but they are broader than synthetic profiles obtained by radiative transfer codes \citet{Ruan2018,Ruan2019}. 
The Dopplershifts were determined  as following: we consider the blue and the red points in a  H$\alpha$ profile distant of 0.6 \AA\ and having the same intensity. The shift of the  middle of the wavelength interval compared to the rest value represents the Dopplershift of the pixel. The zero velocity was obtained by assuming that the sum of all the velocities in the prominence is null. Doppler shift maps are obtained every 30 seconds (see an example in Fig.~\ref{figs:figs:msdp}b) and in the Appendix~\ref{sec:appendix} (Fig.~\ref{figs:MSDP_Maps}). Large sections of the prominence are seen to have coherent velocities. This has been previously discussed in earlier works
\citet{Schmieder2010, Ruan2018}).

\subsection{Mask definition for \ion{Mg}{ii} prominence}\label{sec:mask}

To compare the spatio-temporal evolution and physical properties of the prominence viewed in H$\alpha$ and \ion{Mg}{ii}, we used the IRIS raster data and MSDP observations.
We created a mask to remove the solar disk and the background due to scattered light and noise.
First, we defined the position of the solar limb.
Based on the FWHM map of \ion{Mg}{II}, we defined a solar surface as the level 0.6 \AA~of FWHM of \ion{Mg}{II} k line.
In Fig.~\ref{figs:iris}, the inner contour line (around $x$ = 10 arcsec) corresponds to the solar surface.
All points with FWHM larger than 0.6 \AA~(the left side from the contour line) belongs to the solar disk.\\
The outer edge of the mask corresponds to the contour of the prominence in the H{$\alpha$} intensity map defined from MSDP observation (Mask-H$\alpha$).

 The mask is obtained directly from H${\alpha}$ profiles in each pixel.
Profile intensities need to be much larger than noise, so a parameter M is defined 
as the difference between the intensity in line-centre
and the average intensity in wings at + 0.6 \AA\ and -0.6 \AA.
When this parameter is too low (here M lower than 10 in camera arbitrary units), the pixel is declared out of the prominence.
This value is small as compared with the M-values
exceeding 250 in the brightest parts of prominence.
However we see that possible fluctuations due to noise from pixel to pixel do not appear in velocity maps (Fig.\ref{figs:figs:msdp}b),
even in the weak parts of the prominence, close to the mask.

%
Therefore, the \ion{Mg}{ii}
prominence is defined as the region between the solar limb (left line in Fig.~\ref{figs:iris}) and the outer edge of prominence (right contour in Fig.~\ref{figs:iris}).

\subsection{Comparison of H$\alpha$ and \ion{Mg}{ii} Dopplershifts and FWHM}
The spectral analysis of the \ion{Mg}{ii} lines shows alternating red and blue enhancements in the wings as one moves along the slit (Fig.~\ref{figs:mg2_spectra}). We note that these enhancements are coherent over a distance of a few arcsec along the slit. The structures along the slit are narrower at the top of the prominence.

The histograms of the Doppler velocities for H$\alpha$ and \ion{Mg}{ii} k show that the Dopplershifts of both lines are of the same order of magnitude, however the Dopplershifts in \ion{Mg}{ii} can reach 30~km s$^{-1}$ in a few pixels (Fig.~\ref{figs:his_dop_vel}).
{ We note that the histogram of the H$\alpha$ Doppler velocities shows more blueshifted points compared to the histogram of the Doppler velocities of the \ion{Mg}{ii} lines which show more redshifted points. 
This difference is intrinsic to the definition of the rest velocities. For H$\alpha$
we compute the rest velocity using the mean value of all the velocities in the prominence. For wavelength calibration of \ion{Mg}{ii} lines we used the photospheric line \ion{Ni}{i} (see Sect.~\ref{sec:rest_velocity}). The \ion{Mg}{ii} histogram has been done for raster 8 which has been obtained during 16 min while the H{$\alpha$} dopplergram is obtained in less than 30 sec.

\begin{figure*}[ht!]
\centering
\includegraphics[scale=0.7]{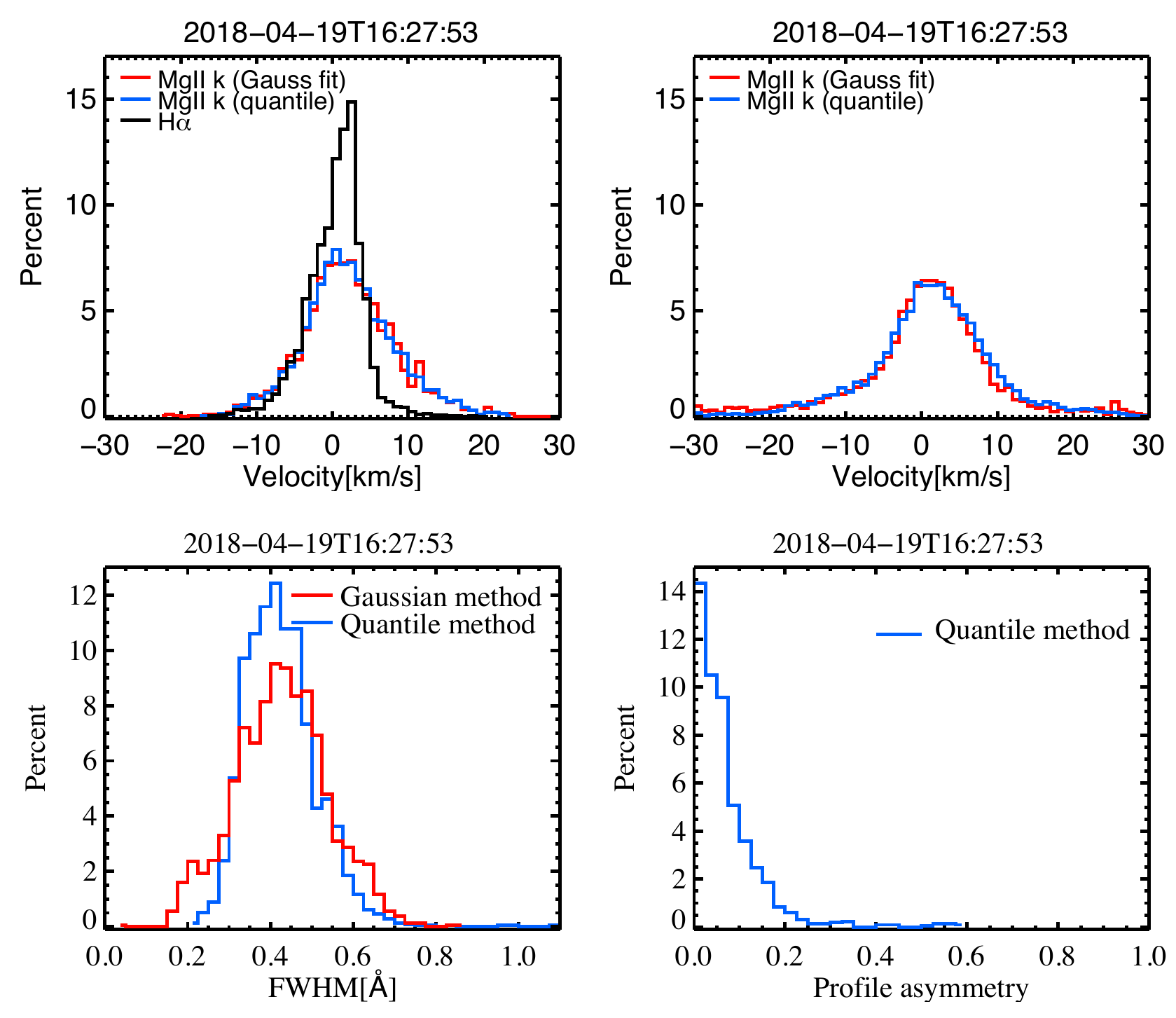}
\caption{ Histograms of Doppler velocities for H$\alpha$ and \ion{Mg}{ii} k lines inside the H$\alpha$ prominence (left top panel) and in the whole \ion{Mg}{ii} prominence (right top panel), FWHM for \ion{Mg}{ii} k line and asymmetry of the \ion{Mg}{ii} k line computed with the quantile method. H$\alpha$ prominence corresponds to pixels enclosed by the contour in Fig.~\ref{figs:figs:msdp} panel a and y-range from 60 to 137 arcsec; The whole \ion{Mg}{ii} k prominence is enclosed in the blue box presented in Fig.~\ref{figs:overview_sji}. The histograms  are based only on the spectral profiles with an intensity larger than the noise. 
\label{figs:his_dop_vel}}
\end{figure*}

 The FWHM of the \ion{Mg}{ii} k line is computed via Gaussian fitting and the quantile method. The two results are in agreement with a FWHM between 0.2 to 1~\AA\ (Fig.~\ref{figs:his_dop_vel}). With the quantile method we show that a fraction of profiles are asymmetric, meaning that their velocities in reality are higher than the computed ones (Fig.~\ref{figs:his_dop_vel}). This effect is also discussed in \citet{Peat2021} (\textit{A\&A submitted}).}

\begin{figure*}[ht!]
\centering
\includegraphics[scale=1.1]{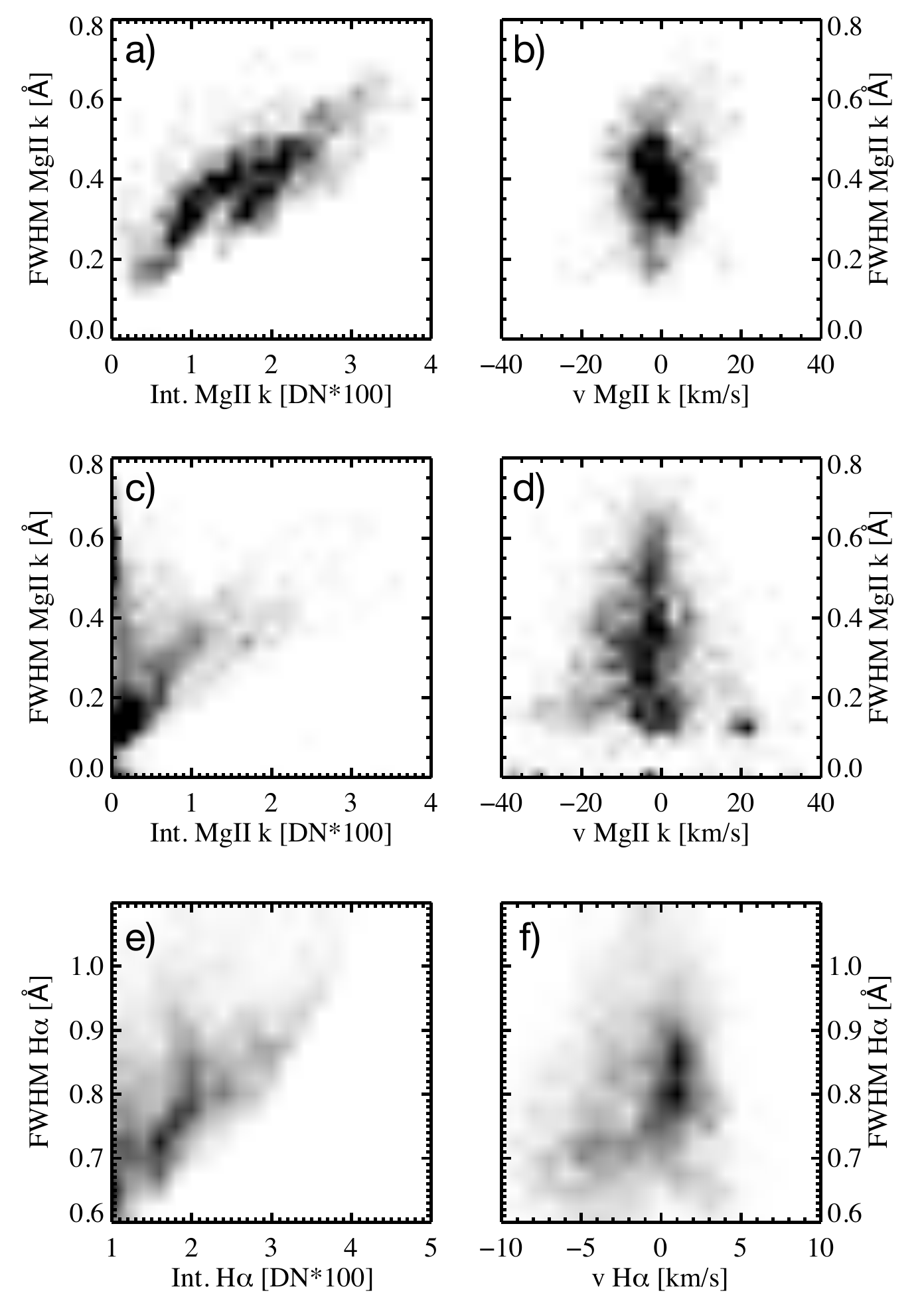}
\caption{Probability density functions (PDFs) for \ion{Mg}{ii} k line parameters for the area between the limb and the outer edge of the H$\alpha$ prominence contour, y-range from 60 to 137 arc in Fig.~\ref{figs:iris}: (a) intensity vs. FWHM, (b) Doppler velocity vs. FWHM; outside of the H$\alpha$ contour (c) intensity vs. FWHM, (d) Doppler velocity vs. FWHM, 
 Probability density functions (PDFs) for H$\alpha$ line parameters (e) intensity vs. FWHM, (f) Doppler velocity vs. FWHM. The PDFs values are plotted in linear scale. 
\label{figs:histoplot_iris_in}}
\end{figure*}

Figure~\ref{figs:histoplot_iris_in} (a, b, c, d) shows the probability density functions for \ion{Mg}{ii} k line in the prominence inside and outside the H$\alpha$ contour respectively.
In Fig.~\ref{figs:histoplot_iris_in} (a), we note that the integrated intensity is roughly proportional to the width of the \ion{Mg}{ii} line profile, with this correlation plateauing for medium width profiles for a large range of integrated intensities. High integrated intensities correspond to wide profiles with a FWHM reaching 0.8~\AA. In Fig.~\ref{figs:histoplot_iris_in} (c) there exists a weak correlation. Weak intensities are observed with profiles of any width but more profiles with a FWHM of 0.2~\AA\ have a weak intensity. For the \ion{Mg}{ii} prominence inside the H$\alpha$ contour, there are no large Dopplershifts. The maximum Dopplershift is around 10 km~s$^{-1}$.
Out of the H$\alpha$ contour the narrow profiles exhibit large Dopplershifts of $\pm$ 25 km~s$^{-1}$.
For H$\alpha$ profiles, there is a correlation between intensity and FWHM. Therefore, relativey narrow profiles (0.6 \AA) have a lower intensity than wider profiles (1 \AA) (Fig.~\ref{figs:histoplot_iris_in} panels (e)). In panel (f) we note that narrow profiles have Dopplershifts between $\pm$5 km~s$^{-1}$, very few points reach a higher velocity. This demonstrates that wide H$\alpha$ profiles are created by the integration of multiple emitting structures with different velocities along the line of sight. This results in wide line profiles. Broad H$\alpha$ profiles are the signature of the integration of many threads along the LOS.

\subsection{Magnetic field configuration}
\begin{figure*}[ht!]
\centering
\includegraphics[width=14cm]{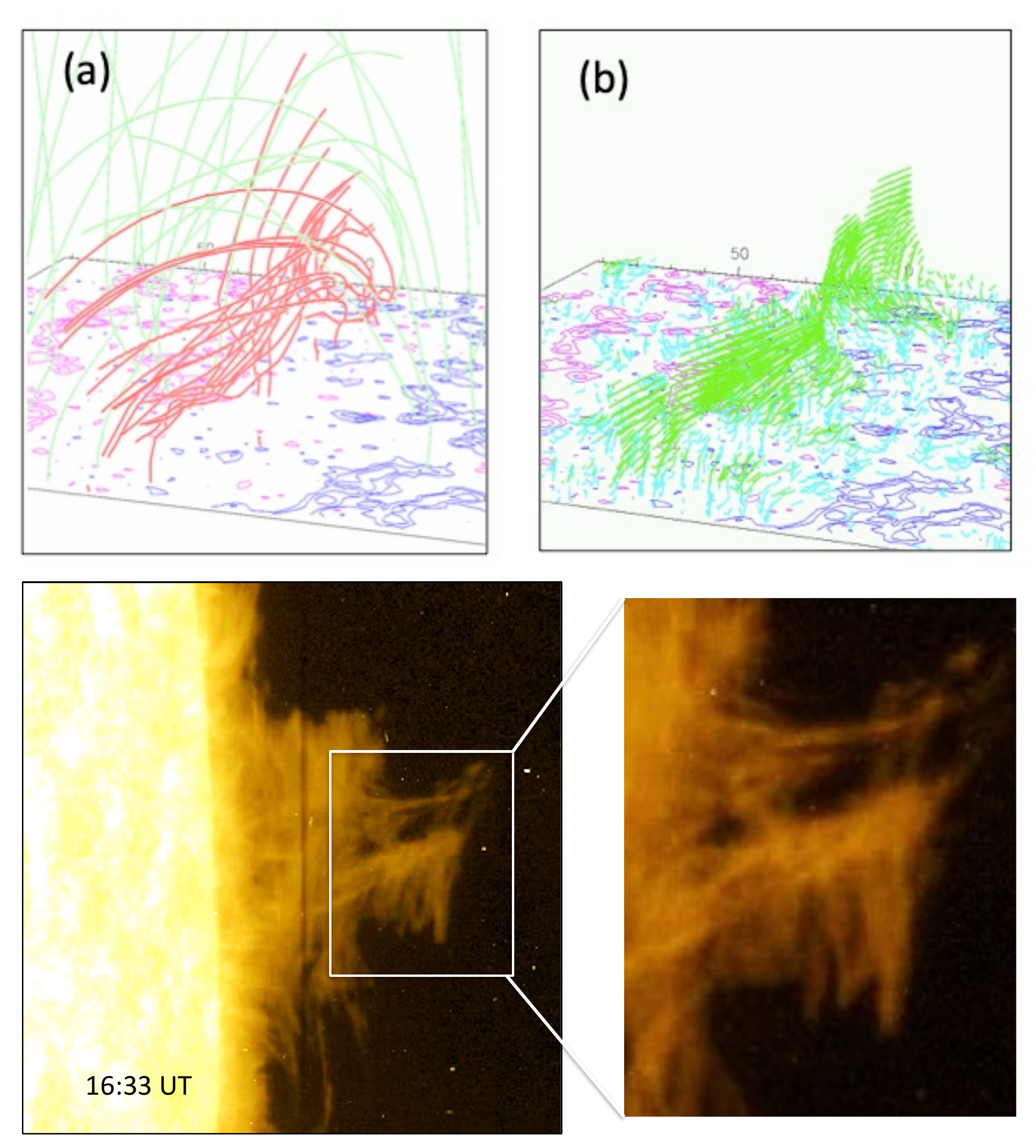}
\caption{ (a) Reconstructed filament with flux rope field lines (red lines) overlaid by arcades (green lines). These field lines are obtained from a LMHD extrapolation of an observed photospheric magnetogram for a different prominence. (b) Dips in the magnetic field lines (green lines). (adapted from \citet{Aulanier2002}). Bottom panels: Zoom on the tornado-like structure showing pile-up of fine horizontal structures. similarly to the green lines in (b).
\label{fig:MHD}}
\end{figure*}

The Dopplershift maps in Fig.~\ref{figs:iris_doppler} clearly show blue/red shift zones parallel to a vertical axis at the top of the prominence (see between 16:11 and 16:27 UT). This suggests twist and rotation. The prominence, when viewed in intensity, has a very classical global morphology with one or two horns at the top (Fig.~\ref{figs:iris_intensity}). However, if we zoom in on the top or look at the IRIS SJI movie we see distinctly horizontal substructures in the top of the vertical column. These are particularly apparent when material escapes from one of the sides of the top of the prominence (Fig.~\ref{fig:MHD} bottom panels). This allows us to conclude that the magnetic configuration of this prominence is similar to classical prominences defined by a flux rope \citep{Aulanier1998,Aulanier2002,vanBallegooijen2004}. Flux rope is commonly represented by twisted magnetic field lines, found by linear force free field extrapolation (Fig.~\ref{fig:MHD} top panels). Filaments are supported by such a flux rope (FR) overlaid by arcades. The idea is that the plasma is located in the dips of the FR and represents the filament viewed in H$\alpha$. The magnetic field lines have been obtained for an other filament as it was crossing the limb by \citet{Aulanier2002} (Fig.~\ref{fig:MHD} left top panel). Above the limb when only the dips are drawn, we see two horns with a pile up of horizontal fine structures like the fine structures in our prominence (right top panel). This kind of interpretation has been confirmed by the models of \citet{Gunar2018} which include radiative transfer. \citet{Gunar2018} could mimic a real prominence viewed from the top and the side with horizontal dips. 

MHD view of a prominence does not consider the plasma in the prominence itself. With IRIS spectroscopy, we explored the characteristics of the plasma loading the dips.

\section{Plasma characteristics from observations and theory}

\label{sec:plasma}
To analyse the plasma characteristics of the prominence using \ion{Mg}{ii} lines we choose two slit positions in the \ion{Mg}{ii} prominence: the main body (slit 12, 16:34 UT) and the top (slit 23, 16:39 UT) representative of the physical state of the prominence (Fig.~\ref{figs:iris} panel a vertical dashed lines). Similar positions are selected in the H$\alpha$ image with the coordinates x= 24 and 46 arcsec (Fig.~\ref{figs:figs:msdp} panel a).

\subsection{\ion{Mg}{ii} integrated intensity, Dopplershifts, FWHM}
\begin{figure*}[ht!]
\centering
\includegraphics[width=14cm]{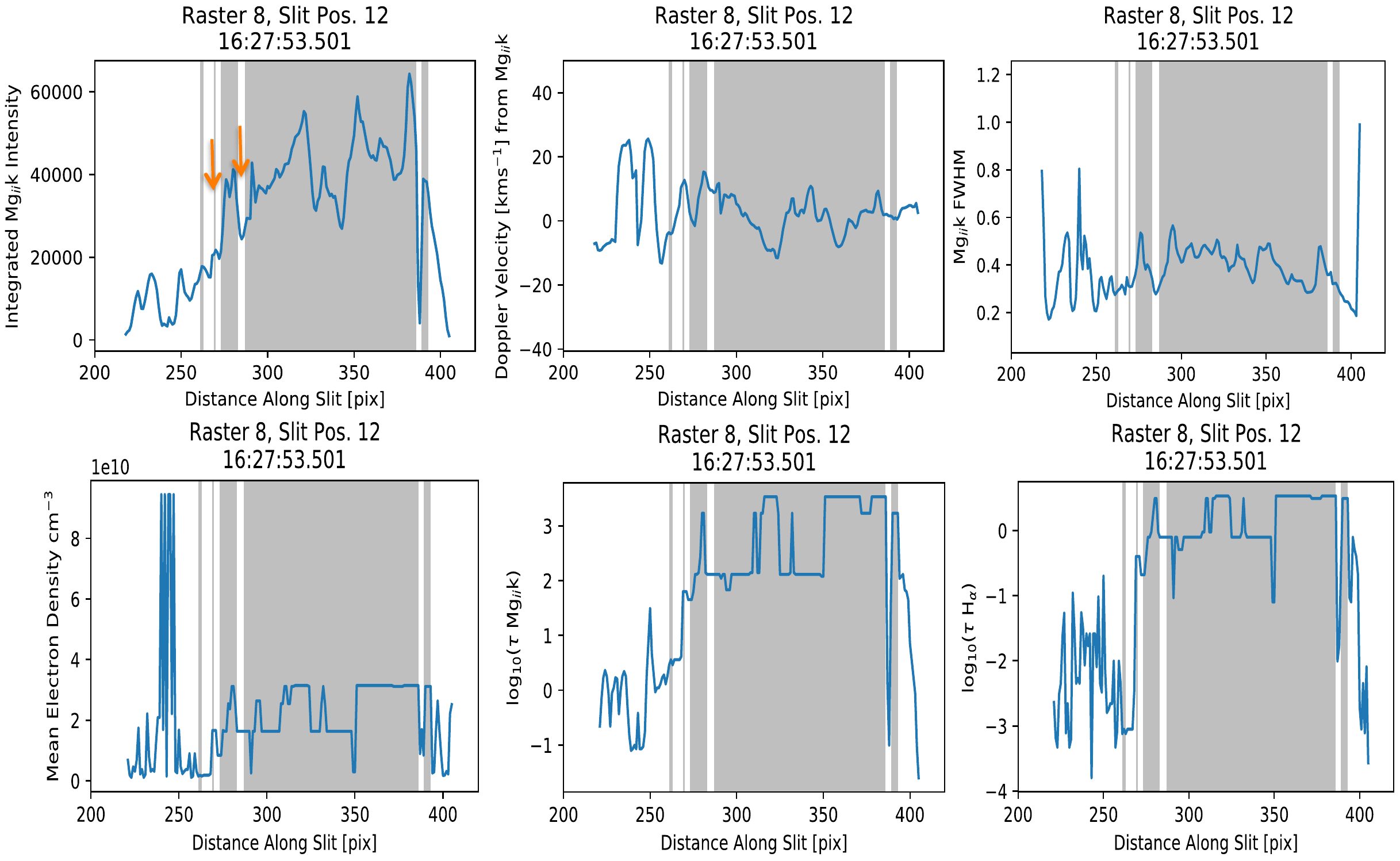}
\caption{Prominence body characteristics: ({\it top panels from left to right}) \ion{Mg}{ii} k observed integrated intensity, Dopplershift, FWHM along slit 12, ({\it bottom panels from left to right}) NLTE model results along slit 12: mean electron density, $ \tau$(\ion{Mg}{ii} k), 
$ \tau$(H $\alpha$). The grey areas correspond to pixels where the observed profiles in \ion{Mg}{ii} cannot be fit correctly with the grid of profiles derived by the NLTE radiative transfer code. The red arrows indicate the points which are in Table~\ref{tab:mg2_ha_sel}.
\label{figs:slide_8_12}}
\end{figure*}

\begin{figure*}[ht!]
\centering
\includegraphics[width=14cm]{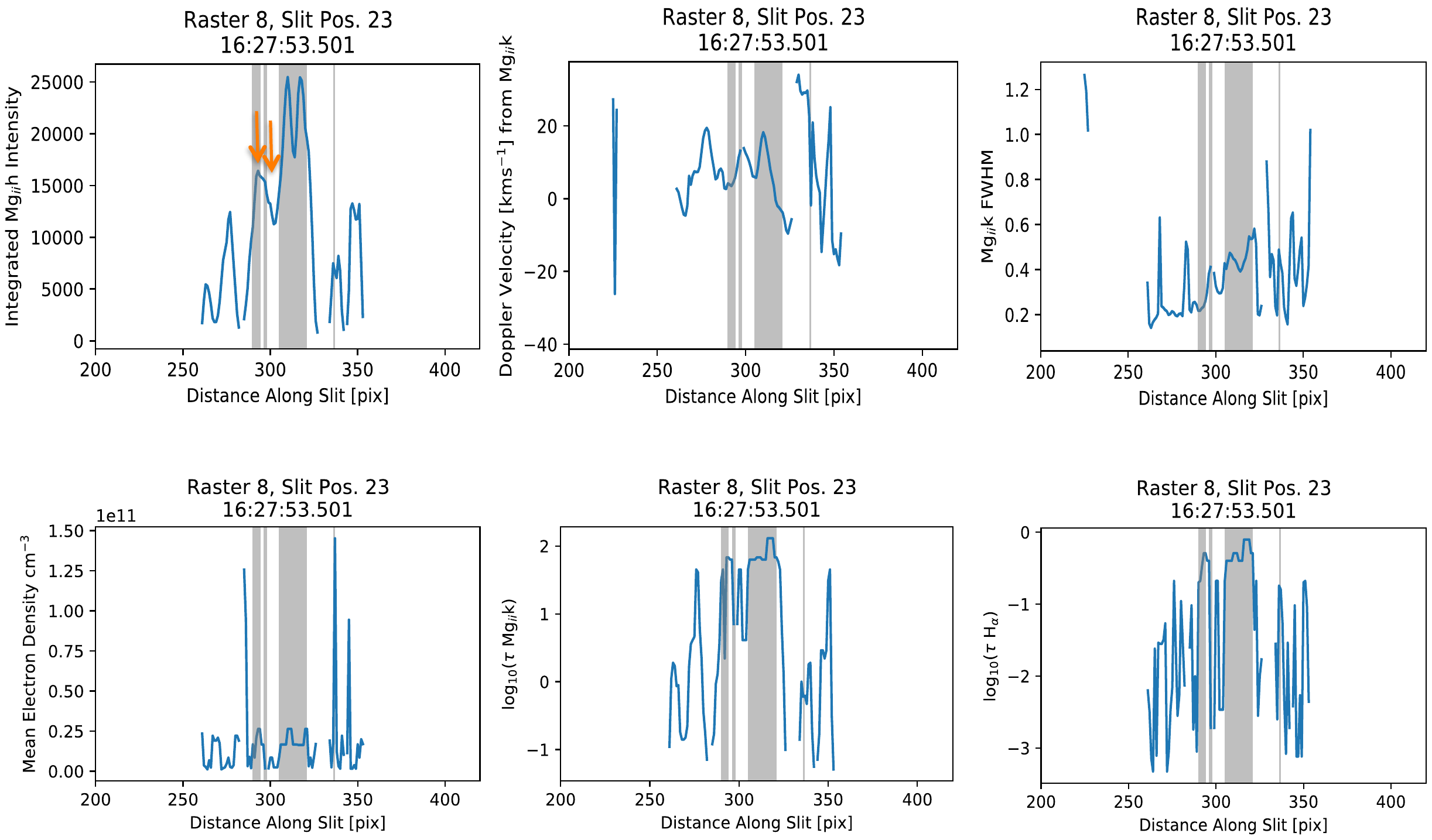}
\caption{ Prominence tornado characteristics: ({\it top panels from left to right}) observed integrated intensity, Dopplershift, FWHM along slit 12, ({\it bottom panels from left to right}) NLTE model results along slit 12: mean electron density, $ \tau$(\ion{Mg}{ii} k), 
$ \tau$(H $\alpha$). The grey areas correspond to pixels where the
observed profiles in \ion{Mg}{ii} cannot be fit correctly with the grid of profiles derived by the NLTE radiative transfer code. The red arrows indicated the points which are in Table~\ref{tab:mg2_ha_sel}.
\label{figs:slide_8_23}}
\end{figure*}

Figures~\ref{figs:slide_8_12} and \ref{figs:slide_8_23} (a, b, c) show the integrated intensity in erg s$^{-1}$
cm$^{-2}$ sr$^{-1}$; the Dopplershifts in km s$^{-1}$; and the FWHM in \AA\, respectively for the \ion{Mg}{ii} k line along the slits 12 and 23 respectively. The prominence is located along the y-axis between pixels 250 and 400 which corresponds to a width of around 50 arcsec. at a height of 20 arcsec. Structures of high intensity are observed along the slit reaching 5 to $6 \times 10^4$~erg s$^{-1}$
cm$^{-2}$ sr$^{-1}$ while the width of the top along slit 23 is narrower, around 33 arcsec. with a mean maximum integrated intensity of $2 \times 10^4$~erg s$^{-1}$
cm$^{-2}$ sr$^{-1}$. 
The raster spectra in Fig.~\ref{figs:mg2_spectra} shows pile-ups of narrow structures (around 2000~km) with the smallest reaching one or two pixels (dimension of the order of 250~km to 500~km).
In Figs.~\ref{figs:slide_8_12} and \ref{figs:slide_8_23} the Dopplershifts range between -5 to 15~km s$^{-1}$ with three main structures of 25 to 35 pixels (8 to 10 arcsec.) in the main part of the prominence whereas the top has narrow velocity structures of around 1.5 to 2 arcsec. The FWHM is 0.3~\AA\ at the edge of the prominence but reaches 0.4~\AA\ in the top and 0.5~\AA\ in the body. The Dopplershift structure include many fine structures with coherent velocities as shown in Fig.~\ref{figs:iris_doppler}.

\subsection{\ion{Mg}{ii} and H$\alpha$ profiles}
Along the two slits we choose individual pixels with a step of one arcsec. at the top (slit 23) (B1-B11) where the structures look narrow and a step of 5 arcsec. in the main body (slit 12) (A1-A15). Table~\ref{tab:mg2_ha_comparison} summarises the characteristics of these profiles. This table allows us to compare these quantitative values with the theoretical values of synthetic profiles obtained from the radiative transfer models of \cite{Levens2019}. 

The \ion{Mg}{ii} and H$\alpha$ profiles from the top of the prominence (B1-B11) at 16:25 UT can be seen in Figs.~\ref{figs:spectrum2} and \ref{fig:MSDP_profiles2} (in Appendix~\ref{sec:appendix}). The profiles from the main part of the prominence (A1-A15) can be seen in the Appendix~\ref{sec:appendix} (Figs.~\ref{figs:spectrum1} and \ref{fig:MSDP_profiles1}). The \ion{Mg}{ii} k profile intensity peaks reach 8 to $12 \times 10^{4}$~erg cm$^{-2}$ s$^{-1}$ sr$^{-1}$ \AA$^{-1}$ in the body (points A1 to A15) and 5 to $10 \times 10^{4}$~erg cm$^{-2}$ s$^{-1}$ sr$^{-1}$ \AA$^{-1}$ in the top (points B1-B11). The mean ratio between the integrated intensities of \ion{Mg}{ii} k versus \ion{Mg}{ii} h is around 1.4 for our sample of points (between 1.38 and 1.55 with an exceptional value of 1.76). This value is typical of prominences \citep{Levens2016,Ruan2018}.

Figures~\ref{fig:MSDP_profiles1} and \ref{fig:MSDP_profiles2} show a set of example MSDP prominence profiles at 16:25 UT from the main part (A1-A15) and the top part (B1-B11) respectively.
The profiles are mostly seen to be wide with a FWHM between 0.653 \AA\ and 0.911 \AA\ (see Table~\ref{tab:mg2_ha_comparison})

The profiles at the top are narrower than in the body because the LOS could cross less structures. In the main body the LOS crosses many structures, each of them having different velocities, resulting in the broadening of the line profiles. The H$\alpha$ profiles are all wide between 0.5 to 0.8 \AA. This results in a larger FWHM than those predicted by 1D models. Single structures cannot be resolved with the spatial resolution of the MSDP spectrograph. The H$\alpha$ line is usually considered as an optically thin line, meaning that many structures are integrated along the LOS \citep{Wiik1992}. However, the H$\alpha$ line may be also optically thick in some cases. The observed FWHM can be larger than this predicted by 1D models for various reasons, especially unresolved structures or optically thick structures with a high speed gradient.

\subsection{Interpretation with a 1D radiative transfer code}
\label{sec:code}
\citet{Levens2019} developed a one-dimensional non-LTE radiative transfer code to understand how physical conditions inside a prominence slab influence shapes and properties of emergent \ion{Mg}{ii} line profiles. This code includes two types of model atmospheres: isothermal and isobaric; and with a prominence-corona-transition-region (PCTR). Using this code they computed a grid of 1007 models, 755 of which contain a PCTR. \ion{Mg}{ii} spectra were computed to demonstrate the influence of the physical parameters of the prominence on the emergent spectra (e.g temperature and geometric thickness). In particular they obtained good correlations between the \ion{Mg}{ii} k line intensities and the intensities of hydrogen lines, and emission measure.Prominence parameters can be retrieved from comparing these synthetic profiles with observations. \ion{Mg}{ii} is ionised for a temperature higher than $T_\mathrm{max}= 30000$~K. Therefore most prominences emitting in \ion{Mg}{ii} lines are cold (T$<$ T$_{max}$), low-pressure, and optically thick structures. Their results agree with previous studies \citep{Heinzel2014,Heinzel2015,Jejcic2018}.

 \subsection{Correlation between \ion{Mg}{ii} and H$\alpha$ lines and emission measure}\label{sec:corr}
 
\begin{figure*}[ht!]
\centering
\includegraphics[width=20cm,angle=90]{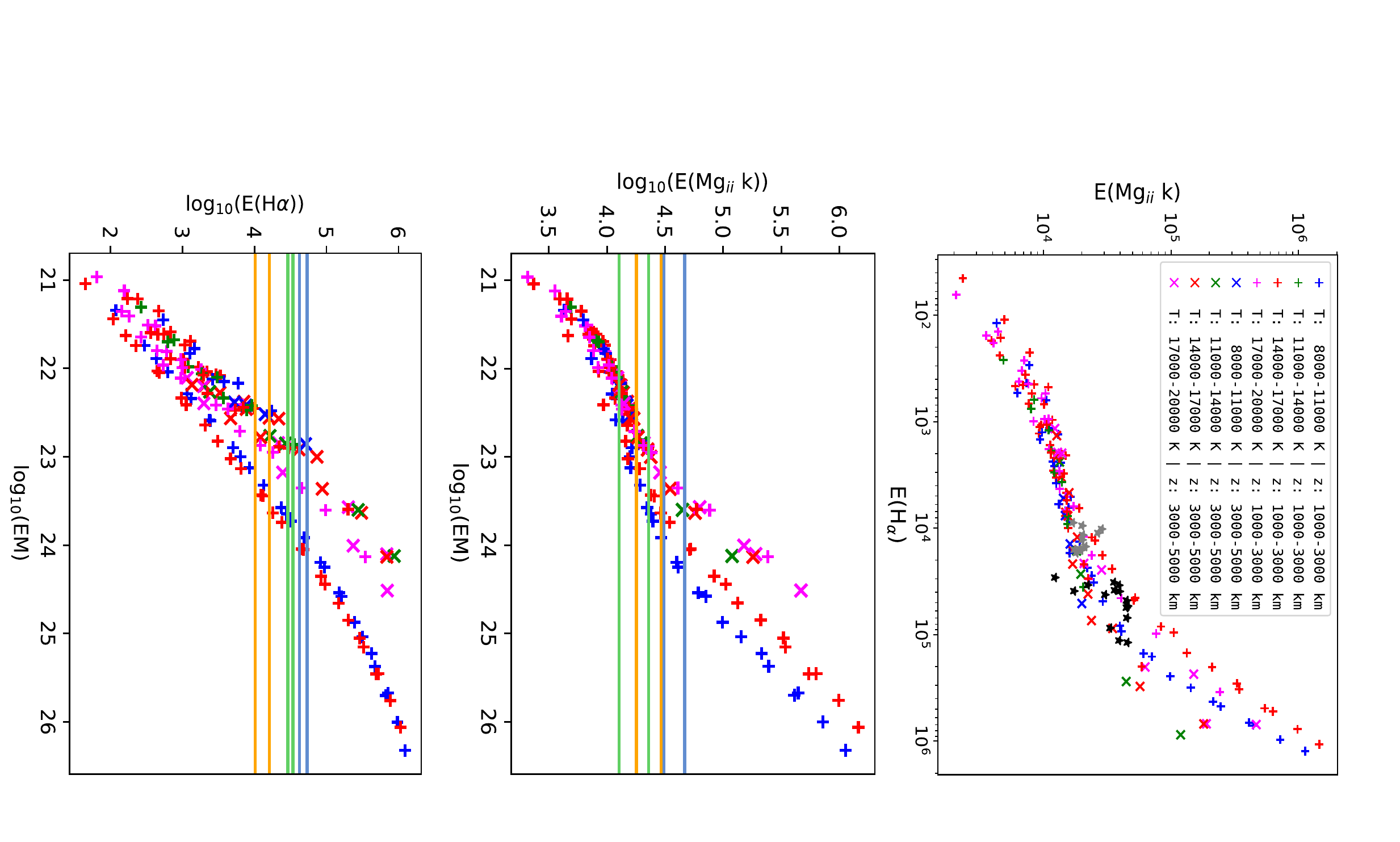}
\caption{Models of radiative transfer for two parameters: temperature between 6000~K and 20000 K, and geometrical thickness between 1000 km and 5000 km. ({\it top panel}): correlation between integrated intensity of \ion{Mg}{ii} k versus H$\alpha$. The black stars represent the observations (see Table~\ref{tab:mg2_ha_comparison}).
({\it middle panel}):
Correlation between integrated intensity and EM for \ion{Mg}{ii} k, ({\it bottom panel}) for H$\alpha$.
The horizontal lines correspond to the extreme values measured in the prominence body (blue), at the edge (green) and in the top (orange).
\label{figs:mgk_em}}
\end{figure*}

 First, we used the global results obtained with the grids of models concerning the integrated intensities of \ion{Mg}{ii} ($E$(\ion{Mg}{ii})) and H$\alpha$ ($E$(H$\alpha$)) and the emission measure (EM). The gas  pressure is an input parameter, so it changes depending on the model. The graphs  in  Fig.~\ref{figs:mgk_em}  contains all 1007 models of the grid. The inputs of the pressure range from 0.1 to 1 dyne cm$^{-2}$. Assumptions on the geometrical thickness and temperature of the structures were made, restricting it to thicknesses of 1000 to 5000 km and temperatures from 6000 K to 20 000 K, with a constant microturbulence equal to 5 km~s$^{-1}$.

\begin{figure*}[ht!]
\centering
\includegraphics[width=18cm]{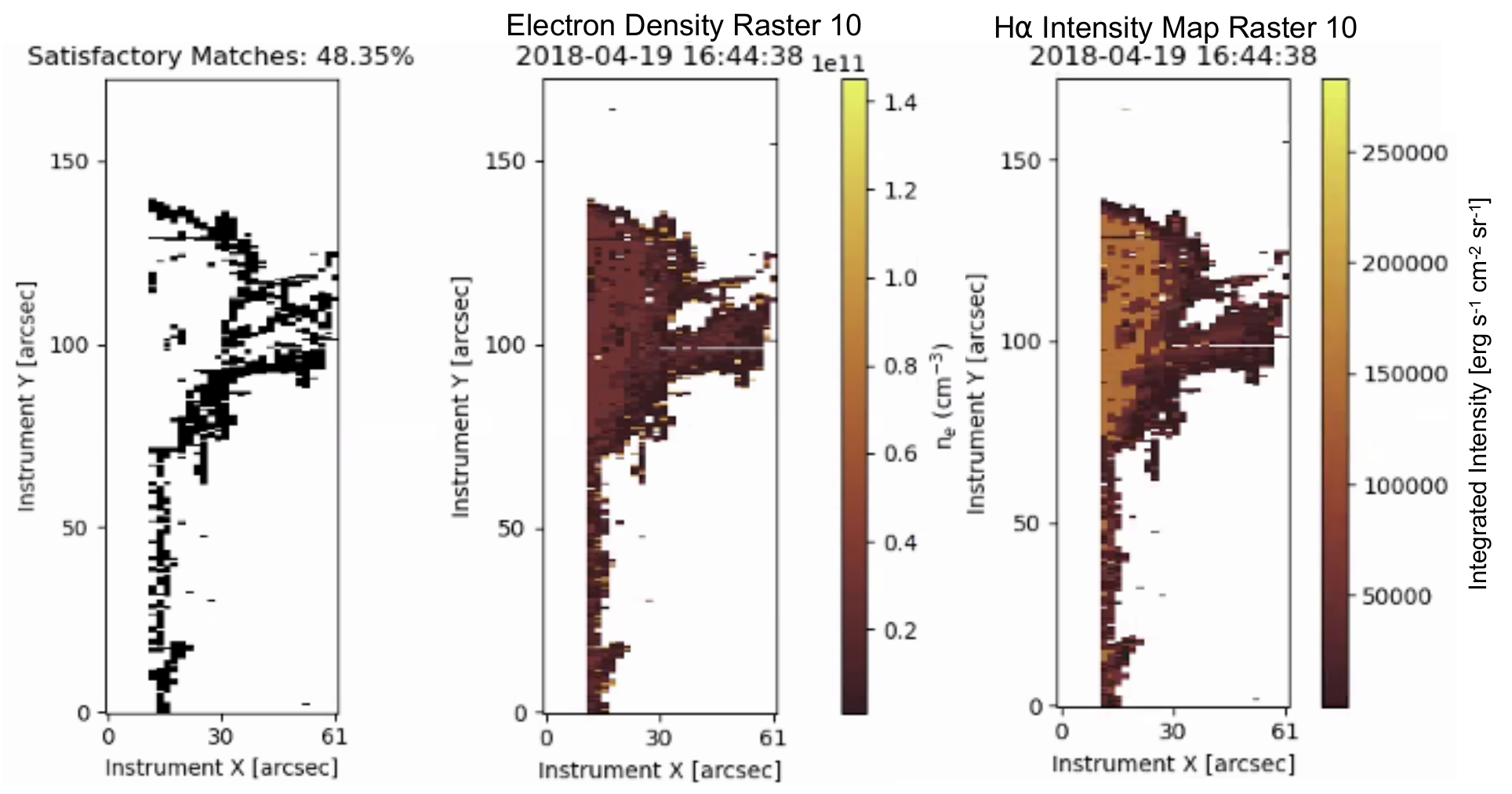}
\caption{ Theoretical results of the NLTE radiative transfer code using the fitting profile method: ({\it left panel}) pixels where the fitting is satisfactory, 
({\it middle panel}) electron density map, ({\it right panel}) intensity in H$\alpha$ according to the models. 
\label{figs:msdp}}
\end{figure*}
 
Figure~\ref{figs:mgk_em} (top panel) shows the variation of $E$(\ion{Mg}{ii}) versus $E$(H$\alpha$) obtained with this sample of theoretical models. The observational points of Table~\ref{tab:mg2_ha_comparison} have been included in the figure.We found a generally good agreement with the theoretical points which means that the co-alignment of the spectra (IRIS and MSDP) have been done correctly. Figure~\ref{figs:mgk_em} (middle and bottom panels) shows the correlation between $E$(\ion{Mg}{ii} k) and EM; and $E$(H$\alpha$) and EM. We can get an approximate value of EM corresponding to the observations. The values of $E$(\ion{Mg}{ii} k), $E$(\ion{Mg}{ii} h) and $E$(H$\alpha$) in Table~\ref{tab:mg2_ha_comparison} concern the top and the main part of the prominence. We separated the main body in two parts, the edge of the prominence, and the central part of the prominence. Two sets of horizontal lines, in green and blue respectively, were drawn bounding the extrema of these regions, whereas the extrema of the top part are bound by orange lines. 
 
From Fig.~\ref{figs:mgk_em} (middle) the edge and top are found to have log(EM) of between 22.1 and 23.8. From Fig.~\ref{figs:mgk_em} (bottom) they are found to have a log(EM) of 22.5 to 23.5. For the body, Fig.~\ref{figs:mgk_em} (middle) gives a log(EM) of 23 to 24.2 and Fig.~\ref{figs:mgk_em} (bottom) gives a log(EM) of 23.5 to 24.4. The agreement between these values is satisfactory. The EM is slightly higher for H$\alpha$ than for \ion{Mg}{ii}.
 
If we estimate that the edge or top of the prominence have a geometrical thickness ($D$) of less than 3000~km, then the values for log(EM) plateau and the uncertainty becomes large for the electron density ($3 \times 10^{9} - 3 \times 10^{10}$ cm$^{-3}$). For the central part of the prominence, where $D$ around 3000~km, the uncertainty is less. The estimate of electron density is around 10$^{10} \pm 0.3 \times 10^{10}$ cm$^{-3}$.
 
\subsection{Observed and synthesized profiles}
We use the recent results of \citet{Peat2021} (\textit{A\&A submitted}) derived from the synthesized profiles obtained with their large data base of models. Their computed line profiles include the two classic shapes of \ion{Mg}{ii}~h\&k profiles (reversed and single peaked). They argue that matching a line profile by comparing the integrated intensities and FWHM of the synthesized profiles to that of the observations is not enough to take into account the two possible kinds of line profiles. They instead compare the computed and observed line profiles point by point in wavelength space.
It is computationally faster to vectorise and fit the whole raster (i.e all of the pixels) at once rather than doing it pixel by pixel. 

At its heart, the procedure is an RMS (root mean square) minimisation routine. The code will always find a fit that it believes to best, regardless of what data it is given. But \citet{Peat2021} (\textit{A\&A submitted}) introduce a test to quantify the level of agreement between the observed profile and the best-fit, computed profile. 
This leads to the use of two filters (applied ad hoc in the routine) -- a prominence filter and a bad fit filter. The former filters out noise and the solar limb, leaving only prominence material, and the latter filters bad fits (i.e. pixels where the agreement between the best modelled profile and the observed profile is not high enough). These two filters have to work together as 'good fits' are found in the noise by the simple metric used by the bad fit filter (an RMS cutoff). The 'best fit' found in the noise is usually a low intensity, single peaked, low FWHM profile. The RMS here is below that of the RMS cutoff because the noise itself is so low that the RMS is low even although the fit is bad.

High intensity profiles are fitted with a mix of double and single peaks, depending on the profile we are fitting to. The best fits are actually from high intensity single peaks and low FWHM. In Fig.~\ref{figs:msdp} (left panel) we present the map of the points where the best fits are obtained. We note that there are at the edge of the \ion{Mg}{ii} prominence.

\subsection{Electron density and optical thickness}
The electron density map has been computed with the Levens and Labrosse radiative transfer code for all the pixels even with a relative low weight for the pixels where the best fit was not found (Fig.~\ref{figs:msdp} middle panel). Using the good co-alignment between the MSDP maps and IRIS rasters, we show an H$\alpha$ intensity map with a superimposed electron density map (Fig.~\ref{figs:msdp} right panel). The red contour limits the H$\alpha$ prominence (hell brown). By comparing maps in right and left panels we note that the H$\alpha$ prominence corresponds nearly to points where the fit of \ion{Mg}{ii} lines was not correct due to their complex profiles.

\begin{figure}[ht!]
\centering
\includegraphics[width=8cm]{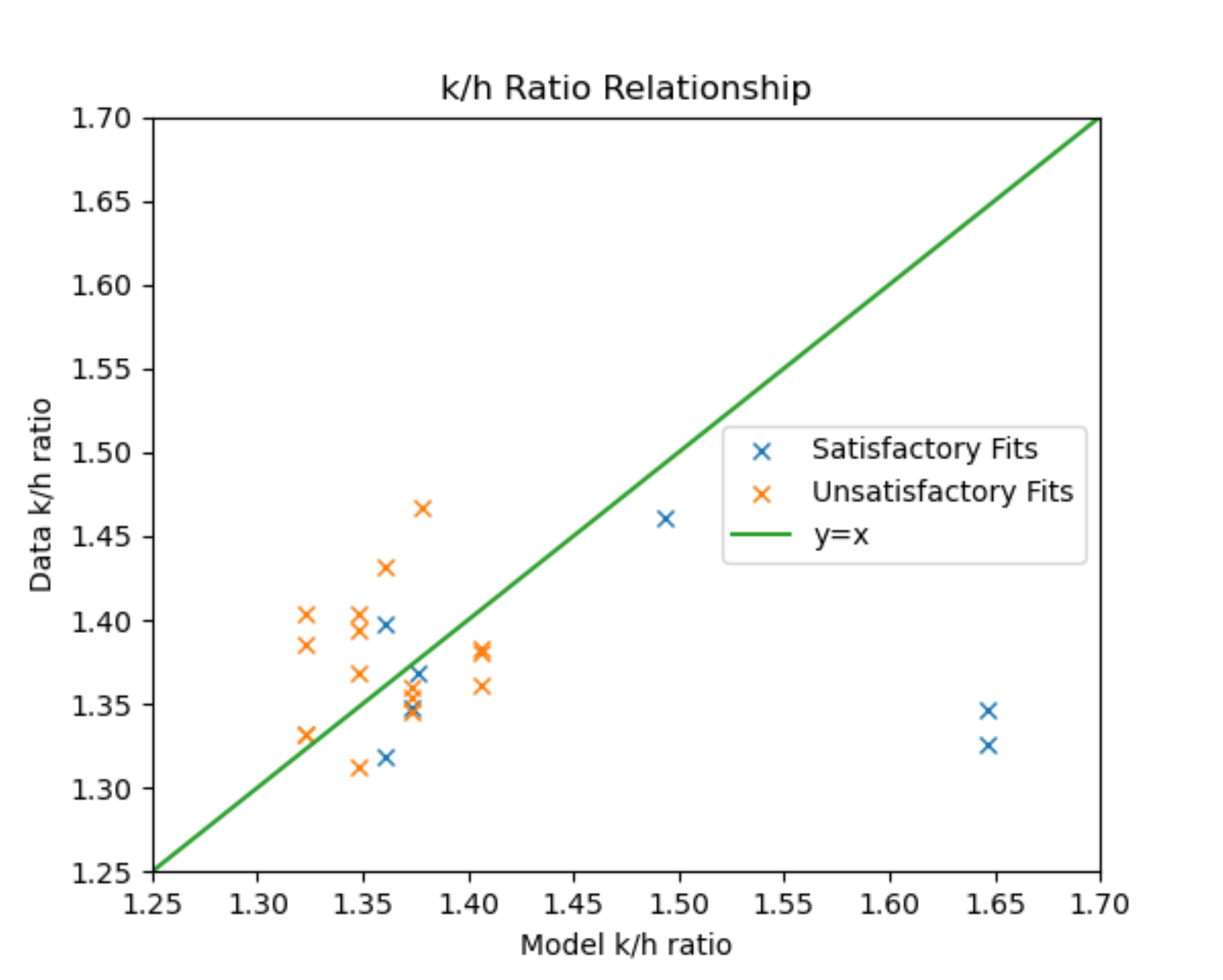}
\caption{Ratio relationship  between the Mg II h and k integrated intensities computed by the   models versus observations (data).
\label{figs:ratio}}
\end{figure}

In order to get parameter quantitative values concerning the prominence and its top 
 we analysed the theoretical results obtained for the two selected slit positions for example mean electron density, optical thickness of \ion{Mg}{ii} k, and optical thickness of H$\alpha$ (Figs.~\ref{figs:slide_8_12} and \ref{figs:slide_8_23} - bottom panels from left to right and Table~\ref{tab:mg2_ha_comparison}). 
 For these pixels we compare the values of the  ratio between the Mg II k and h  integrated intensities  computed with the models and  observed. It is globally in agreement except for some values exceeding typical ratio  values for prominence for example around 1.4
 (Fig.~\ref{figs:ratio}).
Profiles of the  \ion{Mg}{ii} h line show  similar shapes as   \ion{Mg}{ii} k profiles (not shown in the paper).
The gaps in the data are filtered points that the filter found to not be prominence material. In a first step, we analysed the electron density inside the H$\alpha$ prominence, in a second time the pixel in the \ion{Mg}{ii} prominence but out the H$\alpha$ prominence.

 \begin{itemize}
\item Electron density for pixels inside H$\alpha$ prominence

The \ion{Mg}{ii} k profiles of the central part of the prominence have a wide FWHM and a complex profile (Figs.~\ref{figs:spectrum1} and \ref{figs:spectrum2} in Appendix~\ref{sec:appendix}), therefore they could not be fitted by any synthesized profile (Table~\ref{tab:mg2_ha_comparison}).

All synthesized profiles have a FWHM less than 0.3~\AA. As we see in Figs.~\ref{figs:slide_8_12} and \ref{figs:slide_8_23} panels (c) the FWHM exceeds this threshold and is colored in grey. This means that the fit was qualified as 'bad' and the results from the non-LTE radiative transfer code are not reliable in these pixels.
A good fit is mainly possible at the edges and top of the prominence. 
Figure~\ref{figs:msdp} illustrates this well. Due to the ill-matched profiles, we cannot use the code to deduce the electron density in the thick part of the prominence spatially correlated with the H$\alpha$ prominence.
Only the first method based on correlations (Sect.~\ref{sec:corr}) can be used to derive ranges of possible values of the electron density.

With the fitting profile method we retrieved similar results as with the statistical method for the mean electron density. 
However, if we concentrate on a few pixels where the fitting was selected to be 'good' we obtain the following results (see Table~\ref{tab:mg2_ha_sel}).
In the top the electron density 
 is between $1.8 \times 10^{9}$ and $2.66 \times 10^{10}$~cm$^{-3}$. At the edge of the central part of the prominence we have only two values $3.7 \times 10^{9}$ and $1.66 \times 10^{10}$~cm$^{-3}$. The electron density is small (less than 10$^{10}$~cm$^{-3}$) in models with a transition region containing plasma at high temperatures. The profiles obtained in pixels at the top or edge of prominences are fitted either by isothermal or PCTR models.
In the isothermal models, the electron density in the edge of the prominence is higher ($> 10^{10}$~cm$^{-3}$) than in PCTR models.
Points A3 and B2, which are located at the edge of the prominence, have the highest electron density values (1.66 and $2.68 \times 10^{10}$~cm$^{-3}$) compared to points which are closer to the central part of the prominence. For a given geometrical thickness the electron density should commonly decrease to the edge of a prominence. An increase in electron density can be achieved by an enhanced ionisation in this region due to strong irradiation from the coronal environment. The radiation does not fully penetrate into deeper layers of the prominence therefore the ionisation degree and electron density can be lower in the inner parts contrary to previous concepts \citep{Engvold1990}.
\item Electron density for pixels out the H$\alpha$ prominence

The behaviour of increase of the electron density at the top (tornado-like) is confirmed where pixels out of the H$\alpha$ prominence are analysed with the fitting method (Fig.~\ref{figs:slide_8_23} pixel C at x= 345). Such pixels have \ion{Mg}{ii} line profiles with a very low intensity which are well fitted with synthesized profiles obtained with PCTR models leading to very high electron density reaching 10$^{11}$~cm$^{-3}$ and a high temperature larger than 20 000~K. Few pixels as pixel C (shown in Table~\ref{tab:mg2_ha_sel}) have those characteristics. Such higher density compared to the inner prominence density is unusual and will be discussed in the following section. Such possibility was already mentioned in a previous work
\citep{Wiik1992}.
\end{itemize}

The optical thickness of \ion{Mg}{ii} lines have two extreme values so that $\tau_\text{\ion{Mg}{ii} h}$ values is equal  to low  values  around 2 or 3   or high values   between  23 and  32, $\tau_\text{\ion{Mg}{II} k}$ between the double 1.1 and 63 according to the model solution. The low values correspond to pixels at the edge where the plasma is hot (around 23 700~K), while the high $\tau$ values correspond to isothermal models. For PCTR models the temperature is high,   Mg II  becomes a very thin line, this may explain the low optical thickness in that case, otherwise the optical thickness of the order of 30 to 60 is typical in prominences \citep{Levens2016}. In any case the mean pressure is always between 0.01 and 0.05~dyne cm$^{-2}$ which are low values.

\begin{table*}[ht!]
\caption{Characteristics of \ion{Mg}{ii} k and h line profiles of IRIS raster 8 (slit 12 at 16:34:10 UT and slit 23 at 16:39:55 UT) and simultaneous H$\alpha$ line profiles obtained with the MSDP.} \label{tab:mg2_ha_comparison}
\begin{tabular}{@{}lll|cccll|lll@{}}
\hline
\multicolumn{8}{c|}{IRIS (\ion{Mg}{ii})} & \multicolumn{3}{c}{MSDP (H$\alpha$)} \\ \hline
Number &Time & Y& $E$(\ion{Mg}{ii} k) \tablefootmark{a} & $E$(\ion{Mg}{ii} h) \tablefootmark{b} & R(k/h) \tablefootmark{c} &Velocity & FWHM & E(H$\alpha$) \tablefootmark{d} & Velocity & FWHM \\
 & {[}UT{]} &{[}px{]} & & & &{[}km~s$^{-1}${]} & \AA & & {[}km~s$^{-1}${]} & \AA\\
 \hline
A1 & 16:34:10 & 254 & 1.27 & 0.92 & 1.38 & -4.84 & 0.37 & 2.91 & -1.298 & 0.675 \\
A2 & 16:34:10 & 263 & 1.78 & 1.25 & 1.42 & -5.85 & 0.35 & 3.90 & -1.453 & 0.707 \\
A3 & 16:34:10 & 271 & 2.28 & 1.64 & 1.39 & 6.12 & 0.35 & 3.42 & -2.322 & 0.676 \\
A4 & 16:34:10 & 280 & 3.68 & 2.63 & 1.40 & 6.56 & 0.44 & 3.83 & -5.942 & 0.702 \\
A5 & 16:34:10 & 289 & 3.09 & 2.13 & 1.45 & 6.43 & 0.4 & 4.22 & -4.812 & 0.764 \\
A6 & 16:34:10 & 297 & 3.64 & 2.59 & 1.40 & 4.1 & 0.49 & 3.23 & -5.293 & 0.693 \\
A7 & 16:34:10 & 305 & 4.02 & 2.80 & 1.44 & -1.78 & 0.5 & 3.93 & -5.345 & 0.794 \\
A8 & 16:34:10 & 315 & 4.60 & 3.21 & 1.43 & -9.07 & 0.51 & 6.96 & -3.516 & 0.778 \\
A9 & 16:34:10 & 323 & 4.63 & 3.27 & 1.42 & -12.63 & 0.5 & 11.85 & -0.860 & 0.832 \\
A10 & 16:34:10 & 332 & 3.98 & 2.80 & 1.42 & 1.11 & 0.46 & 11.35 & -0.738 & 0.911 \\
A11 & 16:34:10 & 340 & 3.40 & 2.49 & 1.37 & 0.73 & 0.43 & 8.67 & 0.554 & 0.825 \\
A12 & 16:34:10 & 349 & 4.56 & 3.10 & 1.47 & -1.67 & 0.53 & 5.58 & -2.881 & 0.753 \\
A13 & 16:34:10 & 357 & 4.67 & 3.34 & 1.40 & -8.16 & 0.44 & 5.38 & -3.695 & 0.718 \\
A14 & 16:34:10 & 366 & 4.57 & 3.27 & 1.40 & -4.29 & 0.36 & 4.78 & -4.453 & 0.734 \\
A15 & 16:34:10 & 374 & 3.98 & 2.88 & 1.38 & -1.02 & 0.32 & 3.44 & -6.202 & 0.773 \\ \hline
B1 & 16:39:55 & 291	 & 1.85 & 1.25 & 1.48 	 & 0.89	 & 0.24 & 0.89 & -1.469 & 0.688 \\
B2 & 16:39:55 & 294 	 & 2.04 & 1.44 & 1.42	 & 2.41	 & 0.28 & 0.95 & -1.812 & 0.775 \\
B3 & 16:39:55	 & 296	 & 2.07 & 1.45 & 1.42 & 5.57	 & 0.34 & 1.16 & 0.264 & 0.788 \\
B4 & 16:39:55 & 298	 & 2.41 & 1.37 & 1.76	 & 7.44	 & 0.34 & 1.29 & 0.402 & 0.775 \\
B5 & 16:39:55 & 299	 & 2.05 & 1.32 & 1.55	 & 7.92	 & 0.35 & 1.61 & -1.423 & 0.788 \\
B6 & 16:39:55 & 301	 & 1.85 & 1.27 & 1.45	 & 6.77	 & 0.32 & 1.71 & -0.493 & 0.737 \\
B7 & 16:39:55 & 303	 & 1.79 & 1.26 & 1.42	 & 4.62	 & 0.32 & 1.60 & -0.055 & 0.735 \\
B8 & 16:39:55 & 304	 & 1.96 & 1.37 & 1.43	 & 2.72	 & 0.35 & 1.57 & 1.918 & 0.758 \\
B9 & 16:39:55 & 305	 & 2.18 & 1.47 & 1.48	 & 2.57	 & 0.39 & 1.49 & 3.068 & 0.775 \\
B10 & 16:39:55 & 308	 & 2.73 & 1.97 & 1.39	 & 7.64	 & 0.49 & 1.11 & 5.386 & 0.653 \\
B11 & 16:39:55 & 309	 & 2.92 & 2.09 & 1.40	 & 9.63	 & 0.51 & 1.02 & 5.564 & 0.738 \\ \hline
\end{tabular}
\tablefoot{\\
\tablefoottext{a}{Integrated intensity of \ion{Mg}{ii} k line between 2795.49~\AA\ and 2797.53\AA\ in erg sr$^{-1}$ s$^{-1}$ cm$^{-2}$ $\times$ 10$^{4}$}\\
\tablefoottext{b}{Integrated intensity of \ion{Mg}{ii} h line between 2802.57\AA\ and 2804.61\AA\ in erg sr$^{-1}$ s$^{-1}$ cm$^{-2}$ $\times$ 10$^{4}$}\\
\tablefoottext{c}{Ratio between $E$(\ion{Mg}{ii} k) and $E$(\ion{Mg}{ii} h)}\\
\tablefoottext{d}{Integrated intensity of H$\alpha$ line in erg sr$^{-1}$ s$^{-1}$ cm$^{-2}\times$ 10${^4}$.}
}
\end{table*}

\begin{table*}[ht!]
\caption{Plasma parameters of the prominence in raster 8 from NLTE radiative transfer models.}
\label{tab:mg2_ha_sel}
\begin{tabular} {llllllrllll}
\hline
Number &slit & pixel & $Ne$ &$\tau$ (\ion{Mg}{ii} h)& $\tau$ (\ion{Mg}{ii} k)& $T$& $P$ &FWHM k &FWHM h &$\tau$ (H$\alpha$) \\
& & & 10$^{10}$ cm$^{-3}$ & & & K & dyne cm$^{-2}$ & \AA & \AA &\\
\hline
A1& 12& 254 & 0.38 & 0.6 & 1.1 & 23793 & 0.02 & 0.38 & 0.39 & 0.0022 \\
A3& 12& 271 & 1.66 & 32 & 63 & 6000 & 0.05 & 0.37 & 0.34 & 0.40 \\
\hline
B1& 23& 291 & 0.84 & 23 & 45 & 6000 & 0.02 & 0.24 & 0.23 & 0.21 \\
B2& 23 & 294 & 2.68 & 34 & 68 & 6000 & 0.01 & 0.28 & 0.27 & 0.51\\
B5& 23& 299 & 0.18 & 3.6 & 7.1 & 18178 & 0.01 & 0.38 & 0.33 & 0.01 \\
B6& 23& 301 & 0.84 & 23 & 45 & 6000 & 0.02 & 0.33 & 0.31 & 0.21 \\
B7& 23& 303 & 0.23 & 2.1 & 4.1 & 15000 & 0.01 & 0.34 & 0.32 & 0.0034 \\
B8& 23& 304 & 0.23 & 2.1 & 4.1 & 15000 & 0.01 & 0.36 & 0.33 & 0.0034 \\ \hline
C& 23 & 345 & 9.45 & & & 40000& & 0.42 & 0.40 & \\
\hline
\end{tabular}
\end{table*}

\section{Discussion and conclusion}
\label{sec:conclusion}

We present the data concerning a tornado-like prominence observed on April, 19, 2018 with three instruments (SDO/AIA, IRIS, MST/MSDP). The top of the prominence is active and shows twist and rotation in AIA 171 and IRIS SJI 2796 movies, suggesting evidence for a tornado.

We analysed carefully the evolution of the prominence intensity and velocity using IRIS spectra and SJIs. The SJIs show that the top of the prominence consisted of horizontal fine structures where material was going away or towards the sky plane alternatively. We noted the formation of blobs at the top which escape along long field lines with large transverse velocities. We mentioned the similarity of this prominence with previous LFFF magnetic extrapolations obtained for different prominences \citep{Aulanier2002}. In their extrapolation long twisted field lines are part of a flux rope simulating the prominence surrounded by arcades. The prominence is represented by the dips in the field lines. We suggest that the leakeage of the blobs occur along the flux rope magnetic field lines or arcades. With such a configuration it is difficult to infer that the tornado could really turn on itself. Therefore, the Dopplershifts derived from the spectra of IRIS showing adjacent blue and redshifts  in vertical areas at the top  would be interpreted as counter-streaming flows and not rotation.

The second part of the paper concerns the plasma characteristics in the prominence and its top. The results are unexpected and need a discussion.

Plasma parameters have been obtained by using the results of a NLTE radiative transfer code \citep{Levens2019}. We adopted two methods, the first method is using statistics data based on two observables e.g. integrated intensity and FWHM of \ion{Mg}{ii} and H$\alpha$ lines in each pixel, the second method is based on the fitting of \ion{Mg}{ii} h and k line profiles with synthesised profiles using the radiative transfer code. The first method has already been used in previous studies of prominences \citep{Wiik1992,Heinzel2015,Jejcic2018,Ruan2019}. From the theory we found the variation of the integrated intensity of \ion{Mg}{ii} and H$\alpha$ lines according to the emission measure (EM). Then we could estimate the electron density according to assumptions on temperature and geometrical thickness. Electron density increases with the geometrical thickness and decreases with temperature.
With an assumption of narrower geometrical thickness (D), for the low integrated intensity domain (log E(\ion{Mg}{ii} k)=4-4.5), all the curves are overlapping. Therefore we have no precise solutions in the top of the prominence. 

For higher integrated intensities we found an electron density value around $10^{10} \pm 0.3 \times 10^{10}$~cm$^{-3}$. Such intensities are found in the main part of the prominence.

The second method based on fitting profiles was much more powerful, principally for the top of the prominence. We concentrated on the narrower profiles at the top and at the edge of the prominence with low intensity, part which is not visible in H$\alpha$. We found that the electron density in the top could reach 10$^{11}$~cm$^{-3}$. This value is larger than the value for the inner part of the prominence (10$^{10}$~cm$^{-3}$) using prominence-corona transition - region (PCTR) models. We explained this to be due to large ionisation at the edge leading to very high electron density (\citet{Peat2021} (\textit{A\&A submitted})). These points are located at the top of the prominence.

The plasma in these points would be in an ionisation phase due to radiation and this would justify the high electron density with still a high temperature. It could concern the formation by condensation and heating of the plasma in the top \citep{Luna2012,Claes2020}. \citet{Claes2020} found at small scales a strong fragmentation during the formation of prominence plasma which may correspond to the fine structures at the top of our prominence. 
 In a first phase, this material would appear as flowing over the spine of the prominence. This would explain why we see tornadoes only when prominences are passing over the limb. In a second phase, blobs would form and flow along long field lines of the flux rope.
 
 1D NLTE models has limitations for the interpretation of broad profiles which can be the result of double or triple structures along the line of sight. The broad profiles could also be due to high microturbulence. In \citet{Ruan2019} paper, high micro turbulence was used as a proxy of multi structure velocities.
A 2D NLTE radiative model would be suitable to fit more profiles with wide FWHM.

 \bibliographystyle{aa} 
 \bibliography{aa} 

\begin{thebibliography}{60}
\expandafter\ifx\csname natexlab\endcsname\relax\def\natexlab#1{#1}\fi

\bibitem[{Anzer \& Heinzel(2005)}]{Anzer2005}
Anzer, U. \& Heinzel, P. 2005, \apj, 622, 714

\bibitem[{{Aulanier} \& {D{\'e}moulin}(1998)}]{Aulanier1998}
{Aulanier}, G. \& {D{\'e}moulin}, P. 1998, \aap, 329, 1125

\bibitem[{{Aulanier} \& {Schmieder}(2002)}]{Aulanier2002}
{Aulanier}, G. \& {Schmieder}, B. 2002, \aap, 386, 1106

\bibitem[{{Claes} {et~al.}(2020){Claes}, {Keppens}, \& {Xia}}]{Claes2020}
{Claes}, N., {Keppens}, R., \& {Xia}, C. 2020, \aap, 636, A112

\bibitem[{{Culhane} {et~al.}(2007){Culhane}, {Harra}, {Baker}, {van
  Driel-Gesztelyi}, {Sun}, {Doschek}, {Brooks}, {Lundquist}, {Kamio}, {Young},
  \& {Hansteen}}]{Culhane2007}
{Culhane}, L., {Harra}, L.~K., {Baker}, D., {et~al.} 2007, \pasj, 59, S751

\bibitem[{{De Pontieu} {et~al.}(2014){De Pontieu}, {Title}, {Lemen}, {Kushner},
  {Akin}, {Allard}, {Berger}, {Boerner}, {Cheung}, {Chou}, {Drake}, {Duncan},
  {Freeland}, {Heyman}, {Hoffman}, {Hurlburt}, {Lindgren}, {Mathur}, {Rehse},
  {Sabolish}, {Seguin}, {Schrijver}, {Tarbell}, {W{\"u}lser}, {Wolfson},
  {Yanari}, {Mudge}, {Nguyen-Phuc}, {Timmons}, {van Bezooijen}, {Weingrod},
  {Brookner}, {Butcher}, {Dougherty}, {Eder}, {Knagenhjelm}, {Larsen},
  {Mansir}, {Phan}, {Boyle}, {Cheimets}, {DeLuca}, {Golub}, {Gates}, {Hertz},
  {McKillop}, {Park}, {Perry}, {Podgorski}, {Reeves}, {Saar}, {Testa}, {Tian},
  {Weber}, {Dunn}, {Eccles}, {Jaeggli}, {Kankelborg}, {Mashburn}, {Pust},
  {Springer}, {Carvalho}, {Kleint}, {Marmie}, {Mazmanian}, {Pereira}, {Sawyer},
  {Strong}, {Worden}, {Carlsson}, {Hansteen}, {Leenaarts}, {Wiesmann},
  {Aloise}, {Chu}, {Bush}, {Scherrer}, {Brekke}, {Martinez-Sykora}, {Lites},
  {McIntosh}, {Uitenbroek}, {Okamoto}, {Gummin}, {Auker}, {Jerram}, {Pool}, \&
  {Waltham}}]{DePontieu2014}
{De Pontieu}, B., {Title}, A.~M., {Lemen}, J.~R., {et~al.} 2014, \solphys, 289,
  2733

\bibitem[{{Dere} {et~al.}(1997){Dere}, {Landi}, {Mason}, {Monsignori Fossi}, \&
  {Young}}]{Dere1997}
{Dere}, K.~P., {Landi}, E., {Mason}, H.~E., {Monsignori Fossi}, B.~C., \&
  {Young}, P.~R. 1997, \aaps, 125, 149

\bibitem[{{Dud{\'{\i}}k} {et~al.}(2008){Dud{\'{\i}}k}, {Aulanier}, {Schmieder},
  {Bommier}, \& {Roudier}}]{Dudik2008}
{Dud{\'{\i}}k}, J., {Aulanier}, G., {Schmieder}, B., {Bommier}, V., \&
  {Roudier}, T. 2008, \solphys, 248, 29

\bibitem[{{Engvold} {et~al.}(1990){Engvold}, {Hirayama}, {Leroy}, {Priest}, \&
  {Tandberg-Hanssen}}]{Engvold1990}
{Engvold}, O., {Hirayama}, T., {Leroy}, J.~L., {Priest}, E.~R., \&
  {Tandberg-Hanssen}, E. 1990, {Hvar Reference Atmosphere of Quiescent
  Prominences}, Vol. 363 (Springer), 294

\bibitem[{{Golub} {et~al.}(2007){Golub}, {Deluca}, {Austin}, {Bookbinder},
  {Caldwell}, {Cheimets}, {Cirtain}, {Cosmo}, {Reid}, {Sette}, {Weber},
  {Sakao}, {Kano}, {Shibasaki}, {Hara}, {Tsuneta}, {Kumagai}, {Tamura},
  {Shimojo}, {McCracken}, {Carpenter}, {Haight}, {Siler}, {Wright}, {Tucker},
  {Rutledge}, {Barbera}, {Peres}, \& {Varisco}}]{Golub2007}
{Golub}, L., {Deluca}, E., {Austin}, G., {et~al.} 2007, \solphys, 243, 63

\bibitem[{{Gouttebroze} {et~al.}(1993){Gouttebroze}, {Heinzel}, \&
  {Vial}}]{Gouutebroze1993}
{Gouttebroze}, P., {Heinzel}, P., \& {Vial}, J.~C. 1993, \aaps, 99, 513

\bibitem[{{Gun{\'a}r} {et~al.}(2018){Gun{\'a}r}, {Dud{\'\i}k}, {Aulanier},
  {Schmieder}, \& {Heinzel}}]{Gunar2018}
{Gun{\'a}r}, S., {Dud{\'\i}k}, J., {Aulanier}, G., {Schmieder}, B., \&
  {Heinzel}, P. 2018, \apj, 867, 115

\bibitem[{{Gun{\'a}r} {et~al.}(2007){Gun{\'a}r}, {Heinzel}, \&
  {Anzer}}]{Gunar2007}
{Gun{\'a}r}, S., {Heinzel}, P., \& {Anzer}, U. 2007, \aap, 463, 737

\bibitem[{{Gun{\'a}r} {et~al.}(2008){Gun{\'a}r}, {Heinzel}, {Anzer}, \&
  {Schmieder}}]{Gunar2008}
{Gun{\'a}r}, S., {Heinzel}, P., {Anzer}, U., \& {Schmieder}, B. 2008, \aap,
  490, 307

\bibitem[{{Gun{\'a}r} \& {Mackay}(2015)}]{Gunar2015}
{Gun{\'a}r}, S. \& {Mackay}, D.~H. 2015, \apj, 812, 93

\bibitem[{{Gun{\'a}r} \& {Mackay}(2016)}]{Gunar2016}
{Gun{\'a}r}, S. \& {Mackay}, D.~H. 2016, \aap, 592, A60

\bibitem[{{Heinzel} {et~al.}(2015){Heinzel}, {Schmieder}, {Mein}, \&
  {Gun{\'a}r}}]{Heinzel2015}
{Heinzel}, P., {Schmieder}, B., {Mein}, N., \& {Gun{\'a}r}, S. 2015, \apjl,
  800, L13

\bibitem[{{Heinzel} {et~al.}(2014){Heinzel}, {Vial}, \& {Anzer}}]{Heinzel2014}
{Heinzel}, P., {Vial}, J.-C., \& {Anzer}, U. 2014, \aap, 564, A132

\bibitem[{{Jej{\v{c}}i{\v{c}}} {et~al.}(2018){Jej{\v{c}}i{\v{c}}}, {Schwartz},
  {Heinzel}, {Zapi{\'o}r}, \& {Gun{\'a}r}}]{Jejcic2018}
{Jej{\v{c}}i{\v{c}}}, S., {Schwartz}, P., {Heinzel}, P., {Zapi{\'o}r}, M., \&
  {Gun{\'a}r}, S. 2018, \aap, 618, A88

\bibitem[{{Karpen} {et~al.}(2001){Karpen}, {Antiochos}, {Hohensee}, {Klimchuk},
  \& {MacNeice}}]{Karpen2001}
{Karpen}, J.~T., {Antiochos}, S.~K., {Hohensee}, M., {Klimchuk}, J.~A., \&
  {MacNeice}, P.~J. 2001, \apjl, 553, L85

\bibitem[{{Kerr} {et~al.}(2015){Kerr}, {Sim{\~o}es}, {Qiu}, \&
  {Fletcher}}]{Kerr2015}
{Kerr}, G.~S., {Sim{\~o}es}, P.~J.~A., {Qiu}, J., \& {Fletcher}, L. 2015, \aap,
  582, A50

\bibitem[{Labrosse {et~al.}(2010)Labrosse, Heinzel, Vial, Kucera, Parenti,
  Gun\'{a}r, Schmieder, \& Kilper}]{Labrosse2010}
Labrosse, N., Heinzel, P., Vial, J.-C., {et~al.} 2010, Space Sci. Rev., 151,
  243

\bibitem[{{Landi} {et~al.}(2012){Landi}, {Del Zanna}, {Young}, {Dere}, \&
  {Mason}}]{Landi2012}
{Landi}, E., {Del Zanna}, G., {Young}, P.~R., {Dere}, K.~P., \& {Mason}, H.~E.
  2012, \apj, 744, 99

\bibitem[{{Leenaarts} {et~al.}(2013{\natexlab{a}}){Leenaarts}, {Pereira},
  {Carlsson}, {Uitenbroek}, \& {De Pontieu}}]{Leenaarts2013b}
{Leenaarts}, J., {Pereira}, T.~M.~D., {Carlsson}, M., {Uitenbroek}, H., \& {De
  Pontieu}, B. 2013{\natexlab{a}}, \apj, 772, 89

\bibitem[{{Leenaarts} {et~al.}(2013{\natexlab{b}}){Leenaarts}, {Pereira},
  {Carlsson}, {Uitenbroek}, \& {De Pontieu}}]{Leenaarts2013}
{Leenaarts}, J., {Pereira}, T.~M.~D., {Carlsson}, M., {Uitenbroek}, H., \& {De
  Pontieu}, B. 2013{\natexlab{b}}, \apj, 772, 90

\bibitem[{{Lemen} {et~al.}(2012){Lemen}, {Title}, {Akin}, {Boerner}, {Chou},
  {Drake}, {Duncan}, {Edwards}, {Friedlaender}, {Heyman}, {Hurlburt}, {Katz},
  {Kushner}, {Levay}, {Lindgren}, {Mathur}, {McFeaters}, {Mitchell}, {Rehse},
  {Schrijver}, {Springer}, {Stern}, {Tarbell}, {Wuelser}, {Wolfson}, {Yanari},
  {Bookbinder}, {Cheimets}, {Caldwell}, {Deluca}, {Gates}, {Golub}, {Park},
  {Podgorski}, {Bush}, {Scherrer}, {Gummin}, {Smith}, {Auker}, {Jerram},
  {Pool}, {Soufli}, {Windt}, {Beardsley}, {Clapp}, {Lang}, \&
  {Waltham}}]{Lemen2012}
{Lemen}, J.~R., {Title}, A.~M., {Akin}, D.~J., {et~al.} 2012, \solphys, 275, 17

\bibitem[{{Levens} \& {Labrosse}(2019)}]{Levens2019}
{Levens}, P.~J. \& {Labrosse}, N. 2019, \aap, 625, A30

\bibitem[{{Levens} {et~al.}(2016){Levens}, {Schmieder}, {Labrosse}, \&
  {L{\'o}pez Ariste}}]{Levens2016}
{Levens}, P.~J., {Schmieder}, B., {Labrosse}, N., \& {L{\'o}pez Ariste}, A.
  2016, \apj, 818, 31

\bibitem[{{Li} {et~al.}(2016){Li}, {Zhang}, {Peter}, {Priest}, {Chen}, {Guo},
  {Chen}, \& {Mackay}}]{Li2016}
{Li}, L., {Zhang}, J., {Peter}, H., {et~al.} 2016, Nature Physics, 12, 847

\bibitem[{{Luna} {et~al.}(2012){Luna}, {Karpen}, \& {DeVore}}]{Luna2012}
{Luna}, M., {Karpen}, J.~T., \& {DeVore}, C.~R. 2012, \apj, 746, 30

\bibitem[{{Luna} {et~al.}(2015){Luna}, {Moreno-Insertis}, \&
  {Priest}}]{Luna2015}
{Luna}, M., {Moreno-Insertis}, F., \& {Priest}, E. 2015, \apjl, 808, L23

\bibitem[{{Mackay} {et~al.}(2010){Mackay}, {Karpen}, {Ballester}, {Schmieder},
  \& {Aulanier}}]{Mackay2010}
{Mackay}, D.~H., {Karpen}, J.~T., {Ballester}, J.~L., {Schmieder}, B., \&
  {Aulanier}, G. 2010, \ssr, 151, 333

\bibitem[{{Magara}(2007)}]{Magara2007}
{Magara}, T. 2007, \pasj, 59, L51

\bibitem[{{Mein} {et~al.}(2001){Mein}, {Schmieder}, {DeLuca}, {Heinzel},
  {Mein}, {Malherbe}, \& {Staiger}}]{Mein2001}
{Mein}, N., {Schmieder}, B., {DeLuca}, E.~E., {et~al.} 2001, \apj, 556, 438

\bibitem[{{Mein}(1977)}]{Mein1977}
{Mein}, P. 1977, \solphys, 54, 45

\bibitem[{{Mein}(1991)}]{Mein1991}
{Mein}, P. 1991, \aap, 248, 669

\bibitem[{{Okamoto} {et~al.}(2010){Okamoto}, {Tsuneta}, \&
  {Berger}}]{Okamoto2010}
{Okamoto}, T.~J., {Tsuneta}, S., \& {Berger}, T.~E. 2010, \apj, 719, 583

\bibitem[{{Orozco Su{\'a}rez} {et~al.}(2012){Orozco Su{\'a}rez}, {Asensio
  Ramos}, \& {Trujillo Bueno}}]{Orozco2012}
{Orozco Su{\'a}rez}, D., {Asensio Ramos}, A., \& {Trujillo Bueno}, J. 2012,
  \apjl, 761, L25

\bibitem[{Parenti {et~al.}(2012)Parenti, Schmieder, Heinzel, \&
  Golub}]{Parenti2012}
Parenti, S., Schmieder, B., Heinzel, P., \& Golub, L. 2012, ApJ, 754, 66

\bibitem[{{Peat} {et~al.}(2021){Peat}, {Labrosse}, {Schmieder}, \&
  {Barczynski}}]{Peat2021}
{Peat}, A.~W., {Labrosse}, N., {Schmieder}, B., \& {Barczynski}, K. 2021, \aap,
  status: submitted (not yet accepted)

\bibitem[{{Pereira} {et~al.}(2013){Pereira}, {Leenaarts}, {De Pontieu},
  {Carlsson}, \& {Uitenbroek}}]{Pereira2013}
{Pereira}, T.~M.~D., {Leenaarts}, J., {De Pontieu}, B., {Carlsson}, M., \&
  {Uitenbroek}, H. 2013, \apj, 778, 143

\bibitem[{{Pesnell} {et~al.}(2012){Pesnell}, {Thompson}, \&
  {Chamberlin}}]{Pesnell2012}
{Pesnell}, W.~D., {Thompson}, B.~J., \& {Chamberlin}, P.~C. 2012, \solphys,
  275, 3

\bibitem[{{Pettit}(1932)}]{Pettit1932}
{Pettit}, E. 1932, \apj, 76, 9

\bibitem[{{Ruan} {et~al.}(2019){Ruan}, {Jej{\v{c}}i{\v{c}}}, {Schmieder},
  {Mein}, {Mein}, {Heinzel}, {Gun{\'a}r}, \& {Chen}}]{Ruan2019}
{Ruan}, G., {Jej{\v{c}}i{\v{c}}}, S., {Schmieder}, B., {et~al.} 2019, \apj,
  886, 134

\bibitem[{{Ruan} {et~al.}(2018){Ruan}, {Schmieder}, {Mein}, {Mein}, {Labrosse},
  {Gun{\'a}r}, \& {Chen}}]{Ruan2018}
{Ruan}, G., {Schmieder}, B., {Mein}, P., {et~al.} 2018, \apj, 865, 123

\bibitem[{{Scharmer} {et~al.}(2003){Scharmer}, {Bjelksjo}, {Korhonen},
  {Lindberg}, \& {Petterson}}]{Scharmer2003}
{Scharmer}, G.~B., {Bjelksjo}, K., {Korhonen}, T.~K., {Lindberg}, B., \&
  {Petterson}, B. 2003, in Society of Photo-Optical Instrumentation Engineers
  (SPIE) Conference Series, Vol. 4853, Innovative Telescopes and
  Instrumentation for Solar Astrophysics, ed. S.~L. {Keil} \& S.~V. {Avakyan},
  341--350

\bibitem[{{Schmieder} {et~al.}(2010){Schmieder}, {Chandra}, {Berlicki}, \&
  {Mein}}]{Schmieder2010}
{Schmieder}, B., {Chandra}, R., {Berlicki}, A., \& {Mein}, P. 2010, \aap, 514,
  A68

\bibitem[{{Schmieder} {et~al.}(1999){Schmieder}, {Heinzel}, {Vial}, \&
  {Rudawy}}]{Schmieder1999}
{Schmieder}, B., {Heinzel}, P., {Vial}, J.~C., \& {Rudawy}, P. 1999, \solphys,
  189, 109

\bibitem[{{Schmieder} {et~al.}(2004){Schmieder}, {Lin}, {Heinzel}, \&
  {Schwartz}}]{Schmieder2004}
{Schmieder}, B., {Lin}, Y., {Heinzel}, P., \& {Schwartz}, P. 2004, \solphys,
  221, 297

\bibitem[{{Schmieder} {et~al.}(2017{\natexlab{a}}){Schmieder}, {Mein}, {Mein},
  {Levens}, {Labrosse}, \& {Ofman}}]{Schmieder2017}
{Schmieder}, B., {Mein}, P., {Mein}, N., {et~al.} 2017{\natexlab{a}}, \aap,
  597, A109

\bibitem[{{Schmieder} {et~al.}(1991){Schmieder}, {Raadu}, \&
  {Wiik}}]{Schmieder1991}
{Schmieder}, B., {Raadu}, M.~A., \& {Wiik}, J.~E. 1991, \aap, 252, 353

\bibitem[{{Schmieder} {et~al.}(2014){Schmieder}, {Tian}, {Kucera}, {L{\'o}pez
  Ariste}, {Mein}, {Mein}, {Dalmasse}, \& {Golub}}]{Schmieder2014}
{Schmieder}, B., {Tian}, H., {Kucera}, T., {et~al.} 2014, \aap, 569, A85

\bibitem[{{Schmieder} {et~al.}(2017{\natexlab{b}}){Schmieder}, {Zapi{\'o}r},
  {L{\'o}pez Ariste}, {Levens}, {Labrosse}, \& {Gravet}}]{Schmiederz2017}
{Schmieder}, B., {Zapi{\'o}r}, M., {L{\'o}pez Ariste}, A., {et~al.}
  2017{\natexlab{b}}, \aap, 606, A30

\bibitem[{{Su} {et~al.}(2014){Su}, {G{\"o}m{\"o}ry}, {Veronig}, {Temmer},
  {Wang}, {Vanninathan}, {Gan}, \& {Li}}]{Su2014}
{Su}, Y., {G{\"o}m{\"o}ry}, P., {Veronig}, A., {et~al.} 2014, \apjl, 785, L2

\bibitem[{{van Ballegooijen}(2004)}]{vanBallegooijen2004}
{van Ballegooijen}, A.~A. 2004, \apj, 612, 519

\bibitem[{{Wang} {et~al.}(2016){Wang}, {Chen}, {Fu}, {Li}, {Li}, \&
  {Liu}}]{Wang2016}
{Wang}, B., {Chen}, Y., {Fu}, J., {et~al.} 2016, \apjl, 827, L33

\bibitem[{{Wiik} {et~al.}(1992){Wiik}, {Heinzel}, \& {Schmieder}}]{Wiik1992}
{Wiik}, J.~E., {Heinzel}, P., \& {Schmieder}, B. 1992, \aap, 260, 419

\bibitem[{{W{\"u}lser} {et~al.}(2018){W{\"u}lser}, {Jaeggli}, {De Pontieu},
  {Tarbell}, {Boerner}, {Freeland}, {Liu}, {Timmons}, {Brannon}, {Kankelborg},
  {Madsen}, {McKillop}, {Prchlik}, {Saar}, {Schanche}, {Testa}, {Bryans}, \&
  {Wiesmann}}]{Wulser2018}
{W{\"u}lser}, J.~P., {Jaeggli}, S., {De Pontieu}, B., {et~al.} 2018, \solphys,
  293, 149

\bibitem[{{Xia} {et~al.}(2014){Xia}, {Keppens}, {Antolin}, \&
  {Porth}}]{Xia2014}
{Xia}, C., {Keppens}, R., {Antolin}, P., \& {Porth}, O. 2014, \apjl, 792, L38

\bibitem[{{Yang} {et~al.}(2018){Yang}, {Tian}, {Peter}, {Su}, {Samanta},
  {Zhang}, \& {Chen}}]{Yang2018}
{Yang}, Z., {Tian}, H., {Peter}, H., {et~al.} 2018, \apj, 852, 79

\end{thebibliography}

\begin{acknowledgements}
We thank the observation team at the Meudon solar tower for providing the MSDP data: Pierre and Nicole Mein, Daniel Crussaire and Guiping Ruan. We thank Clara Froment for chairing a fruitful discussion of our observations during the COSPAR 2021 session. Figure~\ref{fig:MHD} has been provided by Guillaume Aulanier.
This study benefited from financial support from the Programme National Soleil Terre (PNST) of the CNRS/INSU, as well as from the Programme des Investissements d'Avenir (PIA) supervised by the Agence nationale de la recherche. The work of KB is funded by the LabEx Plas@Par which is driven by Sorbonne Universit\'e. 
AWP acknowledges financial support from the Science and Technology Facilities Council (STFC) via grant ST/S505390/1. NL acknowledges support from STFC grants ST/P000533/1 and ST/T000422/1.
\end{acknowledgements}

\begin{appendix} 
\section{ H$\alpha $ and Mg II line profiles from MSDP and IRIS observations }\label{sec:appendix}
\begin{figure*}[ht!]
\centering
\includegraphics[width=16cm]{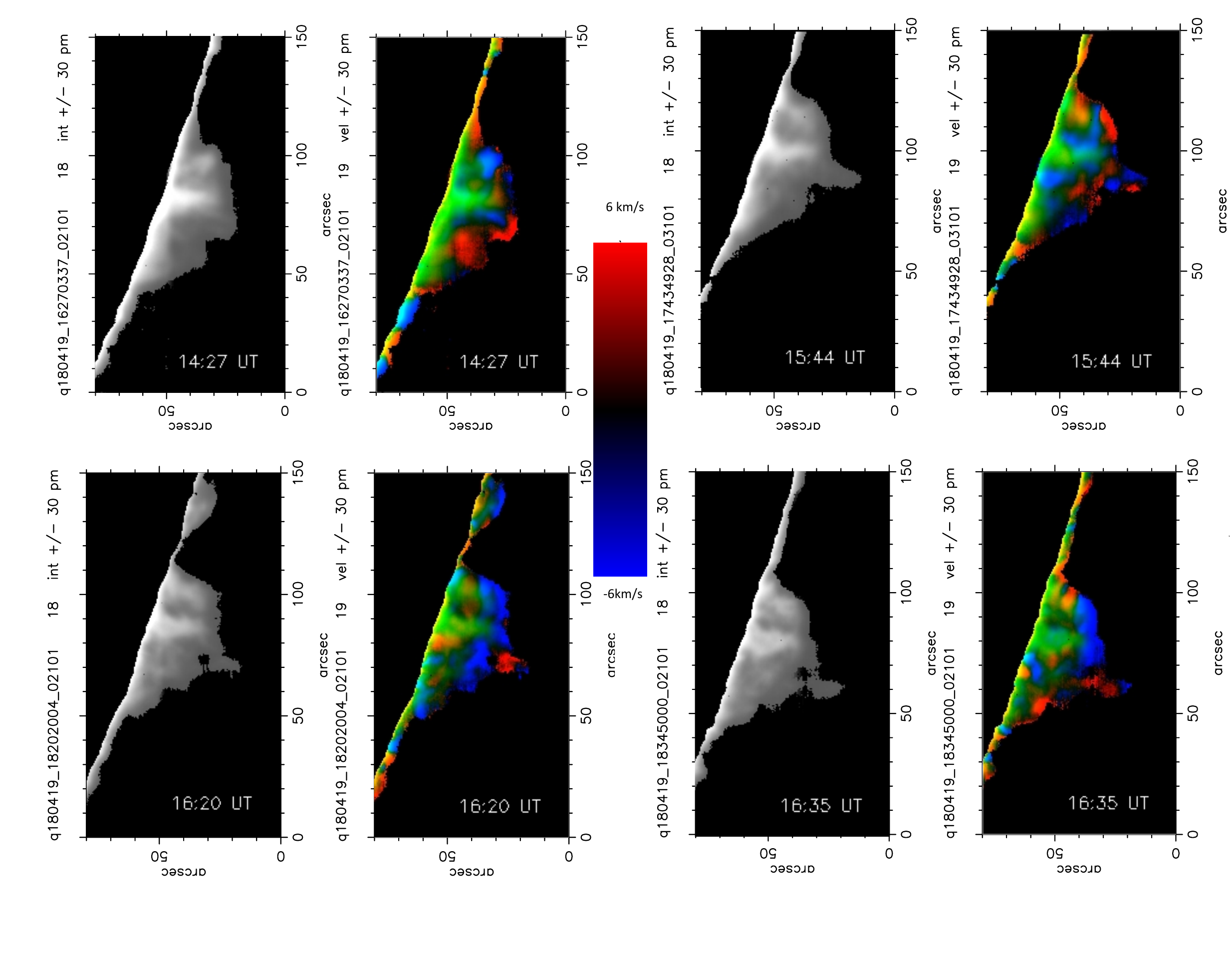}
\caption{MSDP Dopplergrams for different times: intensity and Dopplershifts at H$\alpha \pm 0.3$ \AA. Blue/red are for blue/redshift. Yellow and green are for the intensity. The title on the left of each image indicates the local time (CET) in Meudon. The UT time is indicated in the intensity and Dopplershift maps. }
\label{figs:MSDP_Maps}
\end{figure*}

\begin{figure*}[ht!]
\centering
\includegraphics{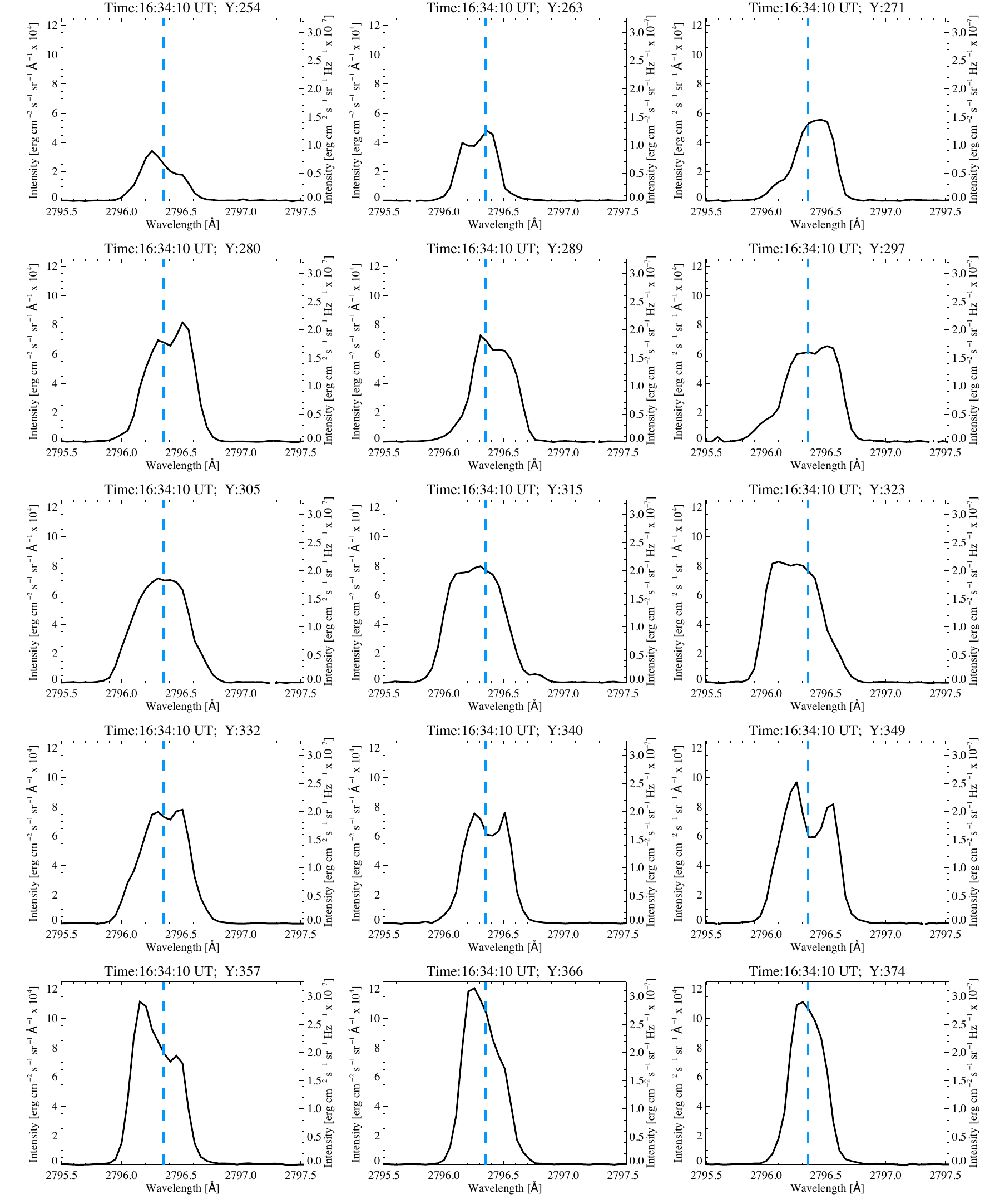}
\caption{IRIS \ion{Mg}{ii} line profiles along slit x=24 arcsec. (slit=12) crossing the prominence at 16:34:10 UT at 15 selected positions indicated in the panel title and corresponding to  pixels along  the vertical left line in Fig.~\ref{figs:iris} (a)  and to Table~\ref{tab:mg2_ha_comparison} (A1-A15). The unit of the $x$-axis is \AA, the units for intensity are indicated on the left in 10$^{4}$~erg cm$^{-2}$ s$^{-1}$ sr$^{-1}$ \AA$^{-1}$, and on the right in 10$^{-7}$~erg cm$^{-2}$ s$^{-1}$ sr$^{-1}$ Hz$^{-1}$.
\label{figs:spectrum1}}
\end{figure*}

\begin{figure*}[ht!]
\centering
\includegraphics{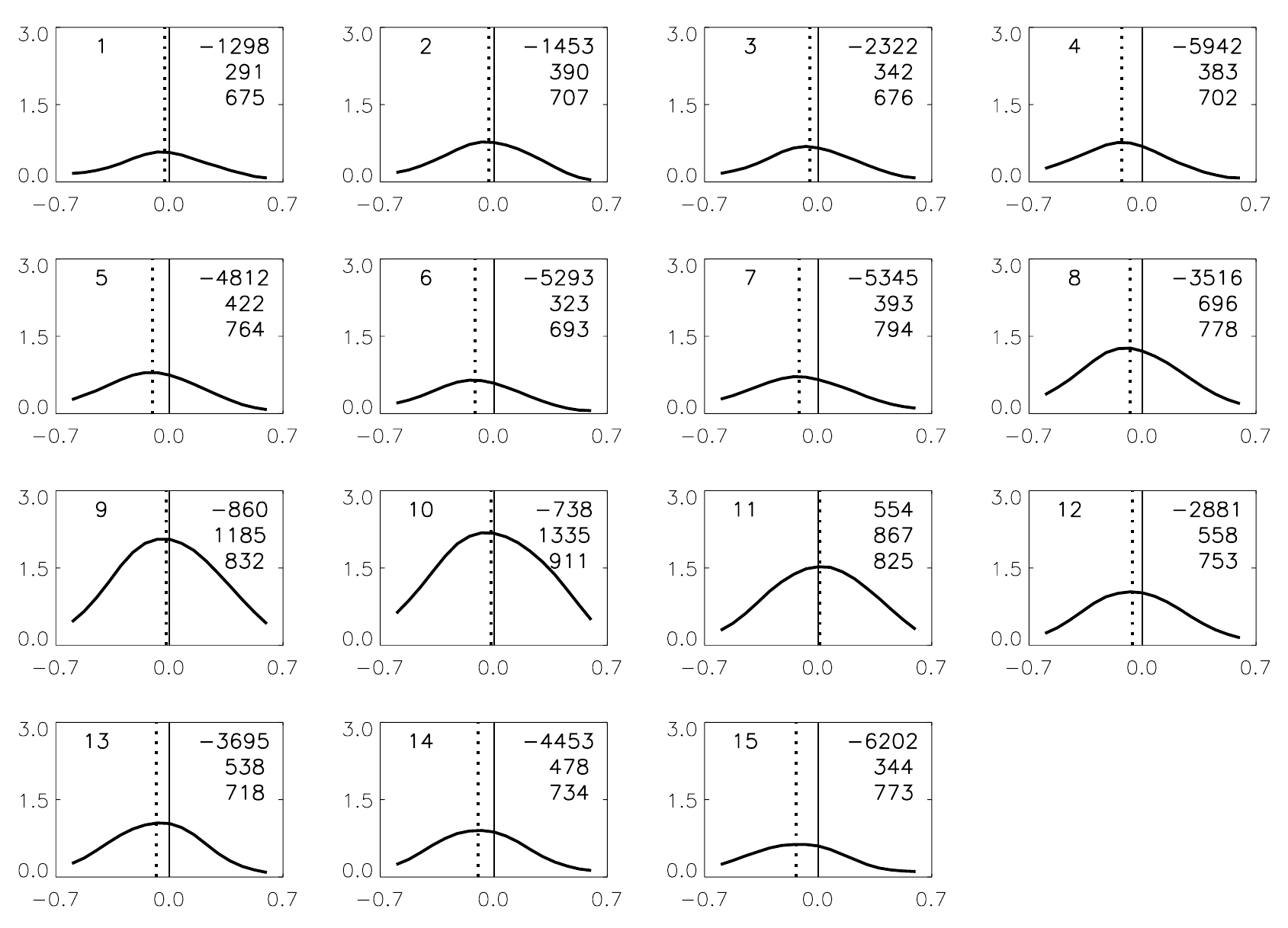}
\caption{Example MSDP prominence profiles in the main part at 16:25 UT. Intensities are in 10$^{-6}$ erg s$^{-1}$ cm$^{-2}$ sr$^{-1}$ Hz$^{-1}$, and wavelength on the $x$-axis is in \AA. In the right corner are indicated the velocity in m/s at H$\alpha$ $\pm$0.3 \AA, below is the integrated intensity in erg sr$^{-1}$ s$^{-1}$ cm$^{-2}$ divided by 100, and on the last raw the FWHM in m\AA. In the left corner, the number of the pixel corresponds to points A1-A15  to  pixels along the vertical A line in Fig.~\ref{figs:figs:msdp} (a)  and in Table~\ref{tab:mg2_ha_comparison}.
\label{fig:MSDP_profiles1}}
\end{figure*}

\begin{figure*}[ht!]
\centering
\includegraphics{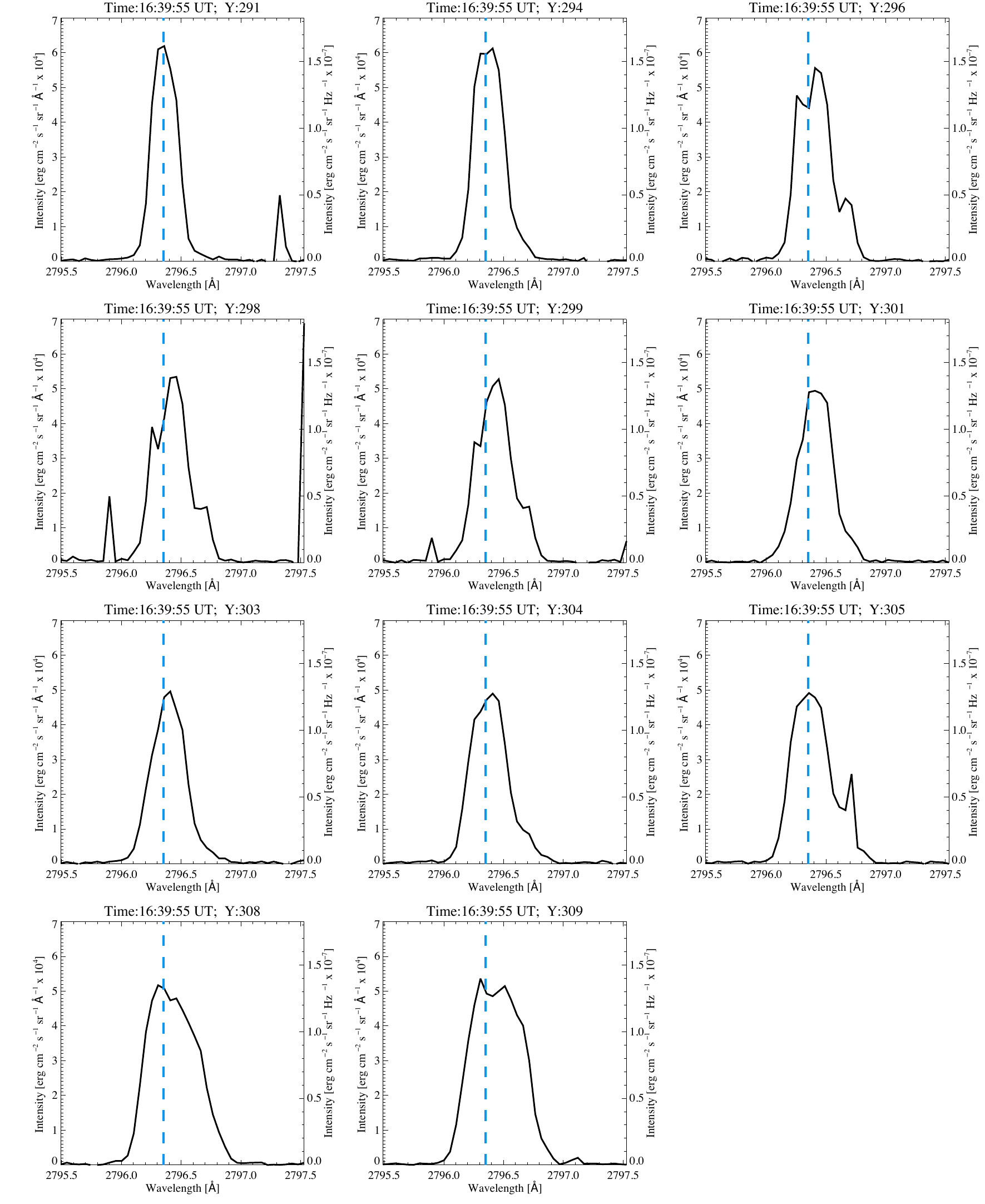}
\caption{IRIS \ion{Mg}{ii} line profiles along slit at $x=46$ arcsec crossing the prominence at 16:39:45 UT in 11 selected positions indicated in the panel title and corresponding to  pixels along the vertical right  line in Fig.~\ref{figs:iris} (a)  and to Table~\ref{tab:mg2_ha_comparison} (B1-B11). The unit of the $x$-axis is \AA, the units for intensity are indicated on the left in 10$^{4}$~erg cm$^{-2}$ s$^{-1}$ sr$^{-1}$ \AA$^{-1}$, and on the right in 10$^{-7}$~erg cm$^{-2}$ s$^{-1}$ sr$^{-1}$ Hz$^{-1}$.
\label{figs:spectrum2}}
\end{figure*}

\begin{figure*}
\centering
\includegraphics{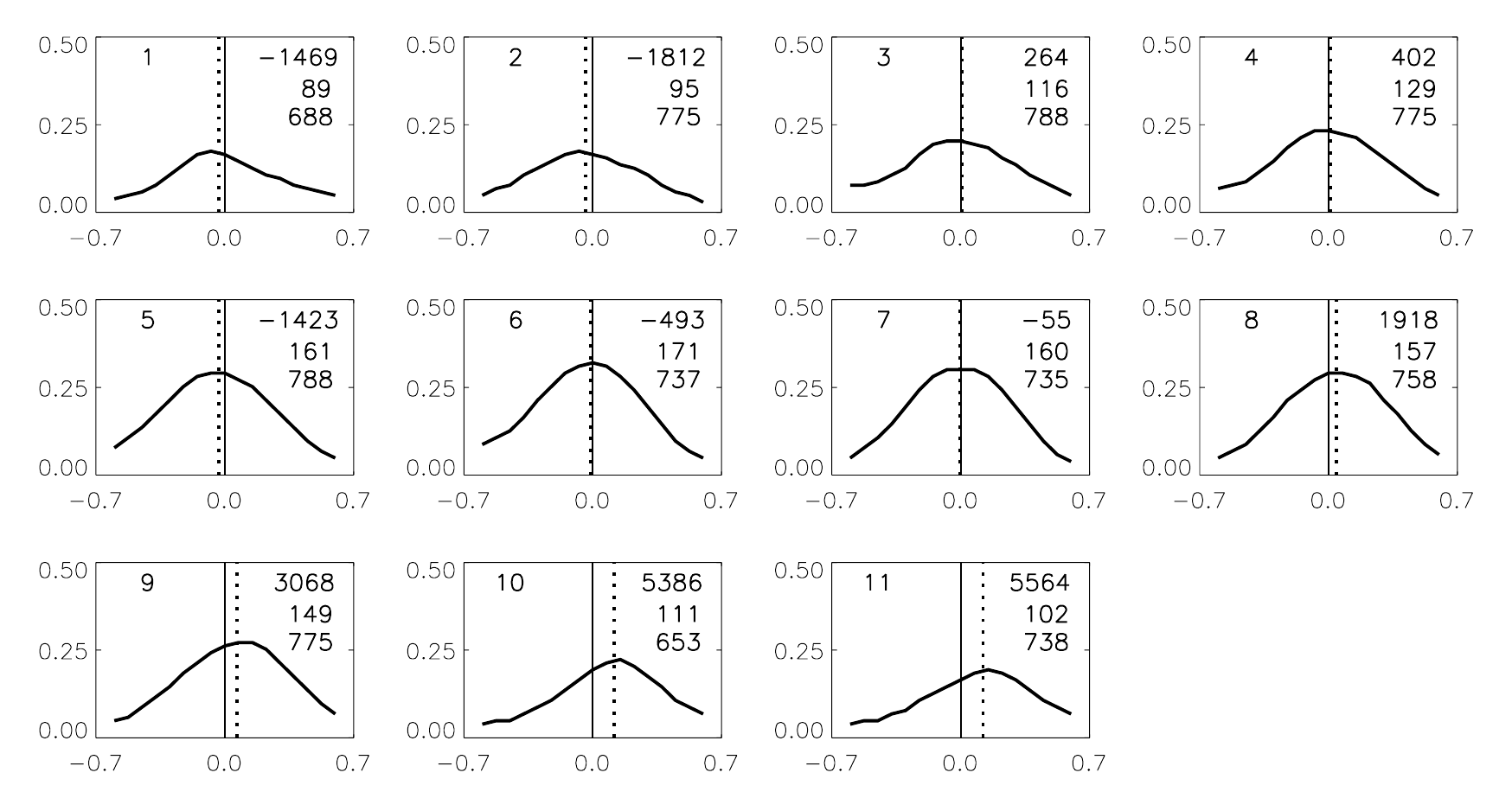}
\caption{Example MSDP prominence profiles in the top at 16:25 UT. Intensities are in 10$^{-6}$~erg s$^{-1}$ cm$^{-2}$ sr$^{-1}$ Hz$^{-1}$, and the unit of the $x$-axis is in \AA. In the right corner are indicated the velocity in m/s at H$\alpha$ $\pm$0.3 \AA, below is the integrated intensity in erg sr$^{-1}$ s$^{-1}$ cm$^{-2}$ divided by 100, and on the last raw the FWHM in m\AA. In the left corner, the number of the pixel at the top corresponds to points B1-B11 in  pixels along the vertical B line in Fig.~\ref{figs:figs:msdp} (a)  and in  Table~\ref{tab:mg2_ha_comparison}.
\label{fig:MSDP_profiles2}}
\end{figure*}

 The MSDP observations cover the time interval of the IRIS observations with some gaps. In the paper we concentrated our study on one time. 
 Here we present four other H$\alpha$ intensity and Doppler maps showing the fast evolution of the velocities versus time (Fig.~\ref{figs:MSDP_Maps}). The
 dopplershifts (blue and red areas) are comparable with the IRIS maps (Fig.~\ref{figs:iris_doppler}). In the paper we show in   Table~\ref{tab:mg2_ha_comparison} the   fitting details of the \ion{Mg}{ii} k and H$\alpha$ line profiles corresponding to IRIS raster 8  slit 12 (the main body of the prominence) and slit 23 in the tornado (top of the prominence).  The correspondence of slits 12 and 23 as lines A (1-15) and B (1-11) are presented  respectively in Figs. \ref{figs:figs:msdp} and \ref{figs:iris} for the two lines.
 Line profiles along the slit 12 (A)  of raster 8 ( main body of the prominence)  are shown  for  \ion{Mg}{ii} k and H$\alpha$ line respectively in Figs.~\ref{figs:spectrum1} and \ref{fig:MSDP_profiles1};  line profiles in the top of prominence (slit 23 -line B) in raster 8  are shown in Figs.~\ref{figs:spectrum2} and \ref{fig:MSDP_profiles2}. IRIS broad profiles with a flat top or reversal could not be fitted with synthesized profiles of the NLTE models.

\end{appendix}

\end{document}